\documentclass[twocolumn]{aastex631}

\usepackage{ae,aecompl}
\usepackage{subfigure}
\usepackage{graphicx}	% Including figure files
\usepackage{amsmath}	% Advanced maths commands
\usepackage{amssymb}	% Extra maths symbols
\usepackage{color,xcolor}
\usepackage{float}
\usepackage{multirow}
\usepackage{makecell}
\usepackage{xparse}
\usepackage{ulem}

\newcommand{\degree}{\hbox{$^\circ$}}

\newcommand{\cm}{\ensuremath{\mbox{~cm}}}

\newcommand{\pcmcu}{\ensuremath{\cm^{-3}}}

\newcommand{\msun}{$M_{\odot}$}

\newcommand{\cmcm}{cm$^{-2}$}
\newcommand{\egcite}{\citep[e.g.,][]}

\newcommand{\hii}{H\textsc{ii}}

\NewDocumentCommand{\fengwei}{ m g }{%
    \IfNoValueTF{#2}
        {\textcolor{teal}{[Fengwei: #1]}}
        {{\color{red}\sout{#1}} \textcolor{teal}{[Fengwei: #2]}}
}

\received{xxx}
\revised{xxx}
\accepted{xxx}

\shorttitle{The ALMA-QUARKS Survey: III}
\shortauthors{Yang et al.}

\begin{document}
\title{\bf The ALMA-QUARKS Survey: III. Clump-to-core fragmentation and search for high-mass starless cores}

\submitjournal{ApJS}
\correspondingauthor{Dongting Yang, Hong-Li Liu, Tie Liu, Xunchuan Liu, Fengwei Xu}
\email{dongting@mail.ynu.edu.cn,hongliliu2012@gmail.com,liutie@shao.ac.cn,liuxunchuan001@gmail.com,fengwei.astro@pku.edu.cn}

\author{Dongting Yang}
\affiliation{School of physics and astronomy, Yunnan University, Kunming, 650091, PR China}
\affiliation{Both authors contributed equally to this work.}

\author{Hong-Li Liu}
\affiliation{School of physics and astronomy, Yunnan University, Kunming, 650091, PR China}
\affiliation{Both authors contributed equally to this work.}

\author{Tie Liu}
\affiliation{Shanghai Astronomical Observatory, Chinese Academy of Sciences, 80 Nandan Road, Shanghai 200030, Peoples Republic of China}

\author{Xunchuan Liu}
\affiliation{Shanghai Astronomical Observatory, Chinese Academy of Sciences, 80 Nandan Road, Shanghai 200030, Peoples Republic of China}

\author[0000-0001-5950-1932]{Fengwei Xu}
\affiliation{Kavli Institute for Astronomy and Astrophysics, Peking University, 5 Yiheyuan Road, Haidian District, Beijing 100871, People's Republic of China}
\affiliation{Department of Astronomy, Peking University, 100871, Beijing, People's Republic of China}

\author{Sheng-Li Qin}
\affiliation{School of physics and astronomy, Yunnan University, Kunming, 650091, PR China}

\author{Anandmayee Tej}
\affiliation{Indian Institute of Space Science and Technology, Thiruvananthapuram 695 547, Kerala, India}

\author{Guido Garay}
\affiliation{Departamento de Astronom\'ia, Universidad de Chile, Casilla 36-D, Santiago, Chile}
\affiliation{Chinese Academy of Sciences South America Center for Astronomy, National Astronomical Observatories, Chinese Academy of Sciences, Beijing, 100101, PR China}

\author{Lei Zhu}
\affiliation{Chinese Academy of Sciences South America Center for Astronomy, National Astronomical Observatories, Chinese Academy of Sciences, Beijing, 100101, PR China}

\author[0000-0002-0786-7307]{Xiaofeng Mai}
\affiliation{Shanghai Astronomical Observatory, Chinese Academy of Sciences, 80 Nandan Road, Shanghai 200030, Peoples Republic of China}
\affiliation{School of Astronomy and Space Sciences, University of
Chinese Academy of Sciences, No. 19A Yuquan Road, Beijing 100049,
People’s Republic of China}

\author{Wenyu Jiao}
\affiliation{Shanghai Astronomical Observatory, Chinese Academy of Sciences, 80 Nandan Road, Shanghai 200030, Peoples Republic of China}

\author{Siju Zhang}
\affiliation{Departamento de Astronom\'ia, Universidad de Chile, Casilla 36-D, Santiago, Chile}

\author{Sami Dib}
\affiliation{Max Planck Institute for Astronomy, Königstuhl 17, D-69117 Heidelberg, Germany}

\author[0000-0003-2300-8200]{Amelia M.\ Stutz}
\affiliation{Departamento de Astronom\'{i}a, Universidad de Concepci\'{o}n,Casilla 160-C, Concepci\'{o}n, Chile}

\author[0000-0002-9569-9234]{Aina Palau}
\affiliation{Instituto de Radioastronom\'ia y Astrof\'isica, Universidad Nacional Aut\'onoma de M\'exico, Antigua Carretera a P\'atzcuaro 8701, Ex-Hda. San Jos\'e de la Huerta, Morelia, 58089, Michoac\'an, M\'exico} 

\author[0000-0002-7125-7685]{Patricio Sanhueza}
\affiliation{Department of Astronomy, School of Science, The University of Tokyo, 7-3-1 Hongo, Bunkyo, Tokyo 113-0033, Japan}

\author{Annie Zavagno}
\affiliation{Aix-Marseille Univ, CNRS, CNES, LAM, 38 rue F. Joliot-Curie, 13013, Marseille, France}
\affiliation{Institut Universitaire de France, Paris, France}

\author{A.Y. Yang}
\affiliation{National Astronomical Observatories, Chinese Academy of Sciences, Beijing, 100101, PR China} 
\affiliation{Key Laboratory of Radio Astronomy and Technology, Chinese Academy of Sciences, A20 Datun Road, Chaoyang District, Beijing, 100101, PR China}

\author{Xindi Tang}
\affiliation{Xinjiang Astronomical Observatory, Chinese Academy of Sciences,
150 Science 1-Stree, Urumqi, Xinjiang 830011, People's Republic of China}

\author{Mengyao Tang}
\affiliation{Institute of Astrophysics, School of Physics and Electronic Science, Chuxiong Normal University, Chuxiong 675000, People's Republic of China}

\author{Yichen Zhang}
\affiliation{Department of Astronomy, Shanghai Jiao Tong University, 800 Dongchuan Rd., Minhang, Shanghai 200240, People's Republic of China }

\author[0000-0002-8586-6721]{Pablo Garc\'ia}
\affiliation{Chinese Academy of Sciences South America Center for Astronomy, National Astronomical Observatories, CAS, Beijing 100101, China}
\affiliation{Instituto de Astronom\'ia, Universidad Cat\'olica del Norte, Av. Angamos 0610, Antofagasta, Chile}

\author[0000-0002-1466-3484]{Tianwei Zhang}
\affiliation{Research Center for Astronomical computing, Zhejiang Laboratory, Hangzhou, China}

\author{Anindya Saha}
\affiliation{Indian Institute of Space Science and Technology, Thiruvananthapuram 695 547, Kerala, India}

\author[0000-0003-1275-5251]{Shanghuo Li}
\affiliation{School of Astronomy and Space Science, Nanjing University, 163 Xianlin Avenue, Nanjing 210023, People’s Republic of China}
\affiliation{Key Laboratory of Modern Astronomy and Astrophysics (Nanjing University), Ministry of Education, Nanjing 210023, People’s Republic of China}

\author{Paul F. Goldsmith}
\affiliation{Jet Propulsion Laboratory, California Institute of Technology, 4800 Oak Grove Drive, Pasadena, CA 91109, USA}

\author{Leonardo Bronfman}
\affiliation{Departamento de Astronom\'ia, Universidad de Chile, Casilla 36-D, Santiago, Chile}

\author{Chang Won Lee}
\affiliation{Korea Astronomy and Space Science Institute, 776 Daedeokdae-ro, Yuseong-gu, Daejeon 34055, Republic of Korea}
\affiliation{University of Science and Technology, Korea (UST), 217 Gajeong-ro, Yuseong-gu, Daejeon 34113, Republic of Korea}

\author{Kotomi Taniguchi}
\affiliation{National Astronomical Observatory of Japan, National Institutes of Natural Sciences, 2-21-1 Osawa, Mitaka, Tokyo 181-8588, Japan}

\author{Swagat Ranjan Das}
\affiliation{Departamento de Astronom\'ia, Universidad de Chile, Casilla 36-D, Santiago, Chile}

\author{Prasanta Gorai}
\affiliation{Rosseland Centre for Solar Physics, University of Oslo, PO Box 1029 Blindern, 0315 Oslo, Norway}
\affiliation{Institute of Theoretical Astrophysics, University of Oslo, PO Box 1029 Blindern, 0315 Oslo, Norway}

\author[0009-0003-6633-525X]{Ariful Hoque}
\affiliation{S. N. Bose National Centre for Basic Sciences, Block-JD, Sector-III, Salt Lake City, Kolkata 700106, India}

\author{Li Chen}
\affiliation{School of physics and astronomy, Yunnan University, Kunming, 650091, PR China}

\author{Zhiping Kou}
\affiliation{Xinjiang Astronomical Observatory, Chinese Academy of Sciences,
150 Science 1-Stree, Urumqi, Xinjiang 830011, People's Republic of China}
\affiliation{School of physics and astronomy, Yunnan University, Kunming, 650091, PR China}

\author{Jianjun Zhou}
\affiliation{Xinjiang Astronomical Observatory, Chinese Academy of Sciences,
150 Science 1-Stree, Urumqi, Xinjiang 830011, People's Republic of China}

\author{Yankun Zhang}
\affiliation{Shanghai Astronomical Observatory, Chinese Academy of Sciences, 80 Nandan Road, Shanghai 200030, Peoples Republic of China}

\author[0000-0002-5310-4212]{L. Viktor T\'oth}
\affiliation{Institute of Physics and Astronomy, E\"otv\"os Lor\`and University, P\'azm\'any P\'eter s\'et\'any 1/A, H-1117 Budapest, Hungary}
\affiliation{Faculty of Science and Technology, University of Debrecen, H-4032 Debrecen, Hungary}

\author{Tapas Baug}
\affiliation{S. N. Bose National Centre for Basic Sciences, Block-JD, Sector-III, Salt Lake City, Kolkata 700106, India}

\author{Xianjin Shen}
\affiliation{School of physics and astronomy, Yunnan University, Kunming, 650091, PR China}

\author{Chuanshou Li}
\affiliation{School of physics and astronomy, Yunnan University, Kunming, 650091, PR China}

\author{Jiahang Zou}
\affiliation{School of physics and astronomy, Yunnan University, Kunming, 650091, PR China}
\affiliation{Shanghai Astronomical Observatory, Chinese Academy of Sciences, 80 Nandan Road, Shanghai 200030, Peoples Republic of China}

\author{Ankan Das}
\affiliation{Institute of Astronomy Space and Earth Science P 177, CIT Road, Scheme 7m, Kolkata 700054}

\author{Hafiz Nazeer}
\affiliation{Indian Institute of Space Science and Technology, Thiruvananthapuram 695 547, Kerala, India}

\author{L. K. Dewangan}
\affiliation{Astronomy $\&$ Astrophysics Division, Physical Research Laboratory, Navrangpura, Ahmedabad 380009, India}

\author[0000-0001-7866-2686]{Jihye Hwang}
\affil{Institute for Advanced Study, Kyushu University, Japan}
\affil{Department of Earth and Planetary Sciences, Faculty of Science, Kyushu University, Nishi-ku, Fukuoka 819-0395, Japan}

\author[0000-0002-9875-7436]{James O. Chibueze}
\affiliation{Department of Mathematical Sciences, University of South Africa, Cnr Christian de Wet Rd and Pioneer Avenue, Florida Park, 1709, Roodepoort, South Africa}
\affiliation{Department of Physics and Astronomy, Faculty of Physical Sciences, University of Nigeria, Carver Building, 1 University Road, Nsukka 410001, Nigeria}

\begin{abstract}

The Querying Underlying mechanisms of massive star formation with ALMA-Resolved gas Kinematics and Structures (QUARKS) survey observed 139 infrared-bright (IR-bright) massive protoclusters at 1.3\,mm wavelength with ALMA. This study investigates clump-to-core fragmentation and searches for candidate high-mass starless cores within IR-bright clumps using combined ALMA 12-m (C-2) and Atacama Compact Array (ACA) 7-m data, providing $\sim$1\arcsec\ ($\sim\rm0.02$~pc at 3.7\,kpc) resolution and $\sim\rm0.6\,mJy\,beam^{-1}$ continuum sensitivity ($\sim 0.3~$\msun\ at 30\,K). We identified 1562 compact cores from 1.3\,mm continuum emission using {\it getsf}. Observed linear core separations ($\lambda_{\rm obs}$) are significantly less than the thermal Jeans length ($\lambda_{\rm J}$), with the $\lambda_{\rm obs}/\lambda_{\rm J}$ ratios peaking at $\sim0.2$. This indicates that thermal Jeans fragmentation has taken place within the IR-bright protocluster clumps studied here. The observed low ratio of $\lambda_{\rm obs}/\lambda_{\rm J}\ll 1$ could be the result of evolving core separation or hierarchical fragmentation. Based on associated signatures of star formation (e.g., outflows and ionized gas), we classified cores into three categories: 127 starless, 971 warm, and 464 evolved cores. Two starless cores have mass exceeding 16\,\msun, and represent high-mass candidates. The scarcity of such candidates suggests that competitive accretion-type models could be more applicable than turbulent core accretion-type models in high-mass star formation within these IR-bright protocluster clumps.

\end{abstract}


\section{Introduction} \label{sec:intro}
High-mass stars (>8\,\msun) play a key role in the universe as primary sources of ionizing radiation, heavy element production, interstellar medium mixing and turbulence, and as important ingredients of galactic structure and evolution
(e.g., \citealt{2006ApJ...638..797D,2007ARA&A..45..481Z,2014prpl.conf....3D}). The formation mechanism of high-mass stars has long been a central topic in astrophysics \citep{2018ARA&A..56...41M}. 

The transition from a clump to cores represents a critical stage in high-mass star formation (e.g., \citealt{2025arXiv250116866B}). Two mainstream theoretical models - the ``turbulent core accretion" model \citep{2003ApJ...585..850M} and the ``competitive accretion" model \citep{2001MNRAS.323..785B} - describe this transition invoking different physical processes. In the turbulent core accretion model, a (few) prestellar core(s) are pre-assembled within a massive clump and supported by strong turbulence and/or magnetic fields. The fast gas accretion from the gas reservoir within the clump during the monolithic-like collapse process determines the final mass of a (few) high-mass star(s).
While in the competitive accretion model, low-mass cores first form from thermal Jeans fragmentation of massive clumps and contiune accreting from beyond their immediate environment, for example through large-scale filamentary accretion flows \citep{2019MNRAS.490.3061V, 2020ApJ...900...82P}. Such a multi-scale dynamical mass accretion scenario facilitates the formation of high-mass stars \egcite{2013A&A...555A.112P,2022MNRAS.510.5009L,2023ApJ...946...22D,2023ApJ...953...40Y,2024RAA....24f5003L,2024AJ....167..228L,2024ApJ...960...76P}.

The validation of both models requires high resolution and sensitive (sub-)millimeter observations to spatially resolve individual dense cores. Recent (sub-)arcsecond observations appear to overwhelmingly support the competitive accretion scenario. For instance, fragmentation of  massive clumps  into clusters of low-mass cores has been commonly observed \citep[e.g.,][]{2019ApJ...886..102S,2019ApJ...886...36S,2025arXiv250305663C}.
In contrast, high-mass prestellar cores - defined as those having $>16$\msun\ within 0.01–0.1~pc and considered as key evidence for the turbulent core accretion model - are rarely detected and are mostly limited to case studies if any (\citealt{2014MNRAS.439.3275W,2015MNRAS.453.3785P, 2017ApJ...849...25L, 2023A&A...675A..53B, 2023A&A...674A..75N, 2024ApJ...961L..35M,2025arXiv250209426V}). This low detection rate implies a very short lifetime for high-mass prestellar cores \citep{2018ARA&A..56...41M}.

Previous searches for high-mass prestellar cores have been dedicated to focusing primarily on infrared dark (IR-dark) clouds (e.g., G11.11-P6-SMA1 reported in \citealt{2014MNRAS.439.3275W};  `dragon cloud'-C2c1a in \citealt{2023A&A...675A..53B};  G34-MM1-E1 in \citealt{2024ApJ...961L..35M}), which are considered to be at the earliest stages of high-mass star formation. However, star formation within a massive clump is not necessarily synchronous; in fact, there may be an age sequence, allowing high-mass prestellar cores to exist even in more evolved clumps where other stars have already begun forming. Recent ALMA statistical studies have shown that dense core masses grow significantly from IR-dark clouds to more evolved IR-bright protostellar clumps \citep{2023MNRAS.522.3719L, 2024ApJS..270....9X}. This suggests that evolved clumps could harbor more massive starless cores. This finding motivates searches for high-mass starless cores not only in IR-dark clouds but also in more evolved IR-bright environments. 
In a sample of 11 IR-bright massive clumps, \citet{2024ApJS..270....9X} have found that the maximum starless core mass could be up to 18.4\,\msun, twice that revealed in IRDCs by \citet{2019ApJ...886..102S}. \citet{2025arXiv250209426V} analyzed the ALMA-IMF data $\sim$2000\,au resolution toward 15 high-mass star-forming regions, reporting 12 candidate high-mass prestellar cores ($>16$\,\msun).  Such efforts could therefore provide critical observational constraints on the turbulent core accretion model, testing whether the low detection rate of high-mass starless cores is due to previous observational biases or to their intrinsic rarity.

The physical drivers of clump fragmentation remain debated, with turbulence and thermal pressure being the primary contenders. Some studies argued for the dominance of turbulence, which was thought to increase the Jeans mass and Jeans length, as observed in several IRDCs \citep{2009ApJ...696..268Z,2011A&A...530A.118P,2013ApJ...773..123S,2014MNRAS.439.3275W,2019ApJ...886..130L,2023ApJ...945...81J,2023MNRAS.526.2278A}. Conversely, high-angular-resolution observations toward other massive clumps with varying evolutionary stages, including IR-dark and IR-bright environments, align with a thermal fragmentation scenario \citep{2015MNRAS.453.3785P,2018A&A...617A.100B,2019ApJ...886...36S,2019ApJ...886..102S,2020ApJ...894L..14L,2022MNRAS.516.1983S,2022A&A...657A..30T,2024ApJ...974...95I,2024ApJ...966..171M}. Additionally, \citet{2024ApJS..270....9X} reported that most observed core separations within evolved IR-bright clumps are significantly less than the predicted thermal Jeans length. This was attributed to a decrease of core separation over time due to persistent global gravitational collapse during evolution \citep{2001MNRAS.323..785B,2018A&A...617A.100B,2023MNRAS.520.2306T}.
Other factors, such as magnetic fields and initial density profiles within massive clumps, have also been explored in the literature, where strong magnetic fields and a concentrated density profile were assumed to effectively reduce fragmentation and thus lead to the formation of more massive cores for high-mass star formation \citep{2011MNRAS.413.2741G,2011ApJ...742L...9C,2011ApJ...729...72P,2014ApJ...785...42P,2016A&A...593L..14F,2021ApJ...912..159P,2022A&A...668A.147H,2025ApJ...980...87S}. 

Thanks to ALMA's high angular resolution and sensitivity, the ALMA Three-millimeter Observations of Massive Star-forming regions (ATOMS; \citealt{2020MNRAS.496.2790L}) survey has observed a sample of 146 massive IR-bright protocluster clumps in band\,3 ($\sim3\,\rm mm$). The sample was selected based on their bright CS ($J=2\!-\!1$) emission ($T_{\rm b} > 2\,\rm K$) from a homogeneous Galactic plane survey of candidate ultra-compact (UC)H{\sc ii}\, regions \citep{1996A&AS..115...81B}. The ATOMS survey was aimed to address several key objectives, including but not limited to 
1)\,establishing a high angular resolution catalog of high-mass star formation dense cores \citep{2020MNRAS.496.2790L,2021MNRAS.505.2801L};  
2)\,investigating the chemistry of hot molecular cores  \citep{2022MNRAS.511.3463Q,2022MNRAS.512.4419P,2025A&A...694A.166C,2025MNRAS.tmp..393K,2025arXiv250406802L};  
3)\,studying UCH{\sc ii} regions (e.g., physical properties and associated feedback mechanisms \citep{2020MNRAS.496.2790L,2021MNRAS.505.2801L,2021MNRAS.508.4639Z,2022MNRAS.510.4998Z,2023MNRAS.520.3245Z,2023MNRAS.520..322Z}; 
and 4)\,characterizing prevailing filamentary structures, and their kinematics and dynamics \citep{2021MNRAS.508.4639Z,2022MNRAS.510.5009L,2022MNRAS.511.3618L,2022MNRAS.514.6038Z,2022MNRAS.516.1983S,2023MNRAS.520.3259X,2023ApJ...953...40Y,2024MNRAS.534.3832D}.

As a follow-up to the ATOMS program, and for advancing the scientific objectives already established therein, the Querying Underlying mechanisms of massive star formation with ALMA-Resolved gas Kinematics and Structures (QUARKS; see Sect.\,\ref{sec:obse} for details on observations) survey was designed with high angular resolution capabilities in ALMA band\,6 ($\sim$1.3\,mm).
This study focuses on clump-to-core fragmentation and searches for candidate high-mass starless cores within IR-bright clumps using combined ALMA 12-m (C-2) and Atacama Compact Array (ACA) 7-m data of the QUARKS survey. 
This paper is organized as follows: Section~\ref{sec:obse} describes the ALMA observations and data reduction for the QUARKS survey.  Section~3 presents the analysis of the TM2+ACA combined data, including the core extraction procedure (Sect.~\ref{sub:sec:res:core_extraction}), dense core classification (Sect.~\ref{sub:sec:res:classification}), physical parameter estimation (Sect.~\ref{sub:sec:res:physicalpara}), analysis of core separation (Sect.~\ref{sub:sec:res:separation}) and thermal Jeans length analysis within  protocluster clumps (Sect.~\ref{sub:sec:res:fragmentation}). Section~\ref{sec:dsicussion} discusses 
fragmentation of IR-bright massive protocluster clumps and search results of candidate high-mass starless cores.
Section~\ref{sec:summary} summarizes the major results.

\begin{table*}[!thb]
\centering
\caption{QUARKS TM2 observations. \label{tab:almaobs}}
\setcellgapes{2pt}
% \makegapedcells
% \linespread{1.2}
\renewcommand{\arraystretch}{1.8} % Default value: 1
\begin{tabular}{ccccccccc}
\hline
\hline
$\rm Group\,ID^a$ & $\rm N_{\rm field}$$^{b}$ & Source & Obs. Date & Min./Max. BL & AR & MRS & \multicolumn{2}{c}{Calibrators}  \\
\cline{8-9}
& & & & (m/m) & (\arcsec) & (\arcsec) & Phase & Bandpass/Flux  \\
(1) & (2) & (3) & (4) & (5) & (6) & (7) & (8)&(9) \\
\hline
1 & 7& 08303-09094 & 2022-12-29 & 15.1/500.6 & 0.7 & 7.3 & J0922-3959 & J1037-2934  \\
2 & 4 &10365-11332& 2024-03-19 & 15.1/313.7 & 1.1 & 11.1 & J1047-6217 & J1107-4449  \\
3 & 15 &12320-13295& 2024-03-22 & 15.1/313.7 & 1.1 & 10.9 & J1254-6111 & J1427-4206  \\
4 & 8 &13471-14212& 2024-03-18 & 15.1/313.7 & 1.1 & 10.9 & J1408-5712 & J1427-4206  \\
\hline
5 & 5 &14382-15290& 2024-03-21 & 15.1/313.7 & 1.1 & 11.1 & J1524-5903 & J1427-4206  \\
 &  && 2024-04-12 & 15.1/313.7 & 1.0 & 10.4 & J1524-5903 & J1427-4206  \\
\hline
6 &  28&15384-16177& 2024-03-22 & 15.1/313.7 & 1.0 & 10.9 & J1603-4904 & J1427-4206  \\
7 & 19 &16272-16489& 2024-03-19 & 15.1/313.7 & 1.1 & 11.1 & J1650-5044 & J1617-5848  \\
8 & 7 &16524-17016& 2024-03-19 & 15.1/313.7 & 1.1 & 11.1 & J1711-3744 & J1617-5848  \\
9 & 15 &17136-17278& 2024-03-22 & 15.1/313.7 & 1.1 & 10.9 & J1720-3552 & J1617-5848  \\
10 & 4 &17439-17455& 2022-04-09 & 15.1/313.7 & 1.1 & 9.5 & J1744-3116 & J1924-2914  \\
11 & 6 &17545-18075& 2024-03-19 & 15.1/313.7 & 1.1 & 11.1 & J1755-2232 & J1924-2914  \\
12 & 13 &18079-18264& 2024-03-19 & 15.1/313.7 & 1.1 & 11.1 & J1832-2039 & J1924-2914  \\
13 & 11 &18290-18341& 2024-03-19 & 15.1/313.7 & 1.1 & 11.1 & J1832-2039 & J1924-2914  \\
14 & 11 &18411-18530& 2022-10-26 & 14.8/312.7 & 1.1 & 9.8 & J1851+0035 & J1924-2914  \\
\hline
15 & 3 &19078-19097& 2024-03-06 & 15.1/313.7 & 1.1 & 10.4 & J1919-2914 & J1924-2914  \\
 & & & 2024-03-19 & 15.1/313.7 & 1.1 & 11.1 & J1919-2914 & J1924-2914  \\
\hline
\end{tabular}{
\begin{flushleft}
$^{(a)}$The QUARKS TM2 observations were arranged into 15 scheduling blocks, which are called `Groups' for short. 
\\
$^{(b)}$ The number of single-pointing fields conducted in each group.
\end{flushleft}
}
\end{table*}

\section{ALMA observations and data reduction}\label{sec:obse}
\subsection{QUARKS sample and ALMA observations}\label{subsec:quarks sample}
The QUARKS survey (PIs: Lei Zhu, Guido Garay, and Tie Liu; Project ID: 2021.1.00095.S; \citealt{2024RAA....24b5009L}) is a follow-up program to the ATOMS survey, offering a higher angular resolution (QUARKS $\sim\rm 0.3$\,\arcsec\, for the best angular resolution versus ATOMS $\sim\rm 2$\arcsec) in ALMA band\,6.
After ruling out 7 protocluster clumps not having enough detectable 3mm continuum emission in the ATOMS survey, we observed in the QUARKS survey  139 IR-bright massive protocluster clumps \citep{2024RAA....24b5009L,2024RAA....24f5011X}.

\begin{figure}[ht!]
    \centering
    \includegraphics[angle=0, width=0.4\textwidth]{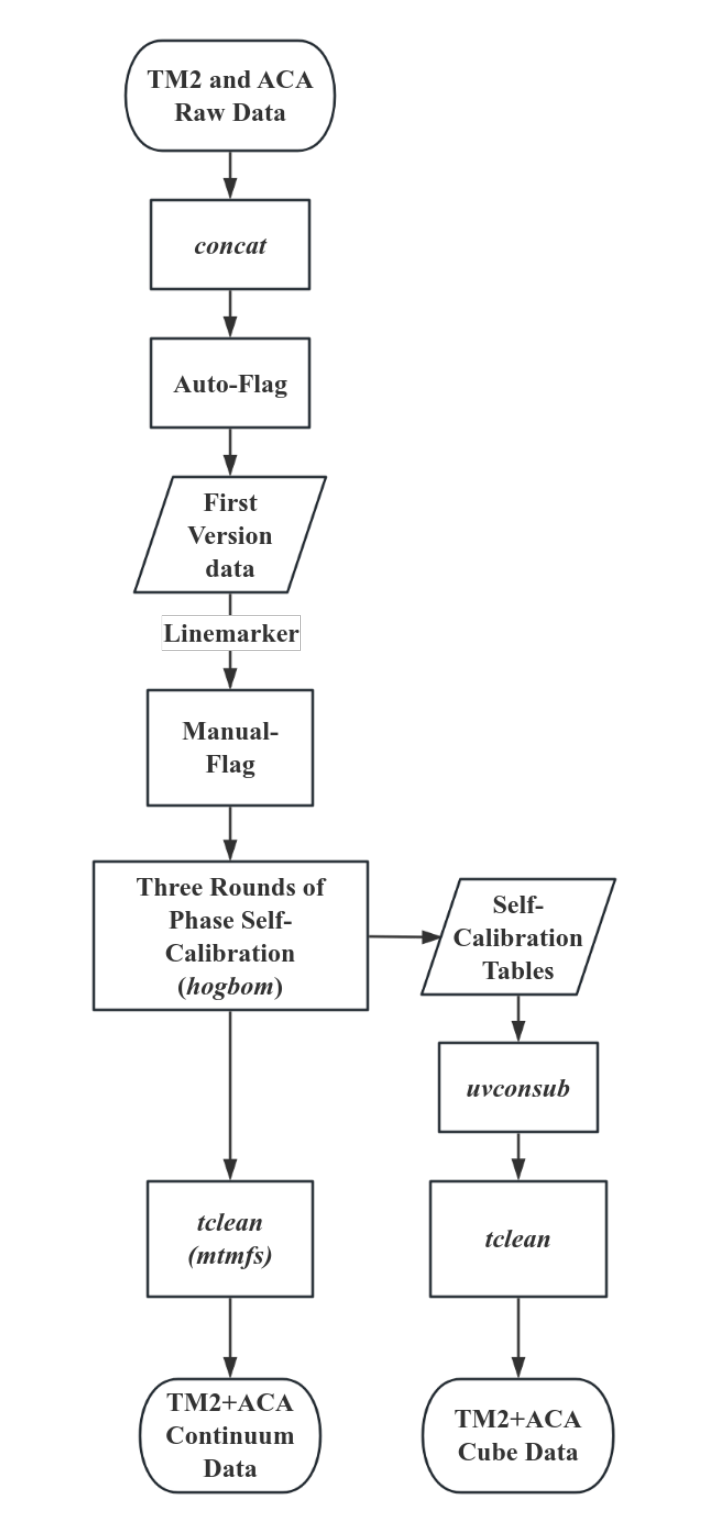} 
    \caption{Flow chart for ALMA-QUARKS ACA and TM2 combined data reduction. 
    The elliptical boxes denote the raw data and the final reduced data. The rectangle indicates a specific processing step, while the parallelogram does an intermediate product during the data reduction.}
    \label{fig:flowchat}    
\end{figure}

Due to the difference in frequency setup between the two surveys,  the field of view (FoV) of the QUARKS only covers a portion of that by the ATOMS (i.e., for the ALMA 12m array, FoV$\sim$20\,\arcsec\ in band 6 for QUARKS versus FoV $\sim$40\,\arcsec\ in band\,3 for ATOMS). Thus, the QUARKS observations were optimized to cover the densest part of the 139 IR-bright protocluster clumps.
To observe extended filamentary structures in 17 of 139 protocluster clumps (e.g., I08448-4343; \citealt{2024ApJ...976..241Y}), two pointing observations were allocated for each of them.
Therefore, the QUARKS survey consists of a total of 156 single-pointings (e.g., fields) to observe 139 protocluster clumps. 
In addition, for each field, the QUARKS observations were obtained with three different ALMA configurations. That is, relatively low ($\sim$5\arcsec), moderate ($\sim$1\arcsec) and high ($\sim$0.3\arcsec) angular resolution observations were conducted by the ACA 7-m array, ALMA 12-m compact array C-2 (TM2) and extended C-5 (TM1) configurations, respectively (see Table\,1 in  \citealt{2024RAA....24b5009L}). ACA observations were first completed in late May 2022 \citep{2024RAA....24f5011X}, followed by TM2 and TM1 that proceeded till June 2024.

Based on similar sky coordinates, the QUARKS 156 fields were grouped into 15 scheduling blocks (SBs), which are referred to as `Group'. 
For TM2 observations, the Group ID, number of fields, and source name within each Group are listed in Columns\,1-3 of Table\,\ref{tab:almaobs}. Note that the Groups are sorted in ascending order by source name.
The date of TM2 observations is listed in Column\,4 of Table\,\ref{tab:almaobs}, starting on April 9, 2022, and completed on March 22, 2024. Note that 2 execution blocks (EBs) were performed on different observing dates for Group\,5 and Group\,15. In the QUARKS TM2 observations, except for the Group\,1, all Groups have similar minimum and maximum baselines (BL; $\sim$ 15.1-313.7\,m), angular resolution (AR; $\sim$1\arcsec), and maximum recoverable scale (MRS; $\sim$ 11\arcsec), which are listed in the Column\,5-7 of Table\,\ref{tab:almaobs}. The phase and bandpass calibrators are listed in Column\,8 and Column\,9 of Table\,\ref{tab:almaobs}, respectively.

The ALMA band\,6 receivers were used in a dual-polarization mode for the QUARKS observations. Four spectral windows (SPWs\,1–4) were configured with a bandwidth of $\sim$2 GHz, a velocity resolution of $\sim\rm 1.3\,km\,s^{-1}$ for each SPW. The frequencies of the four SPWs were centered at approximately 217.92\,GHz, 220.32\,GHz, 231.37\,GHz, and 233.52\,GHz, respectively. The wide range of four SPWs cover numerous major molecular line transitions at $\sim$1.3\,mm, including but not limited to 1)\,cold gas tracer (e.g., $\rm N_2D^+$\,3--2); 2)\,outflow gas tracers (e.g., CO\,(2--1), SiO\,(5--4), $\rm H_2CO$\,(3-2)); 3)\,hot molecular core tracers (e.g., $\rm CH_3CN$\,(12--11), and 4)\,ionized gas tracer, $\rm H30\alpha$. More information about the major molecular lines at 1.3\,mm typical of different gas environments has been summarized in Table\,2 of \cite{2024RAA....24b5009L}.

\subsection{Data Reduction}\label{sub:data}
In this study, the ACA and TM2 observations of the QUARKS survey were combined and imaged using Common Astronomy Software Applications (CASA, version 6.6.0, \citealt{2022PASP..134k4501C}) for both continuum and spectral line emission.
Figure\,\ref{fig:flowchat} presents an overview of the data reduction procedure. 
The {\it Auto-Flag} step was initially conducted to identify line-free channels and subsequently flag data.
The {\it tclean} algorithm was then used to generate the first version of the combined TM2+ACA line data cubes for each field. To optimize the identification of the line-free channels, the {\it Manual-Flag} step was involved iteratively to produce the final continuum (with minimum contamination from line emission) and line data products. Details on data reduction are given below.

\subsubsection{Data Flag and Basic Setups of Data Reduction}\label{subsub:dataset}
As shown in Figure\,\ref{fig:flowchat}, the raw visibility data for the QUARKS ACA and TM2 observations were first concatenated using the CASA task {\it concat}.
In order to execute the CASA task {\it flag}, line-free channels need to be determined for each field. 
For this, we followed the method developed by \cite{2024RAA....24b5009L} for the QUARKS survey. Here, 
1)\,molecular transitions of strong emission lines in the four SPWs were identified as a model spectrum by matching the ALMA pipeline reduced datacube with the laboratory database of spectral lines (CDMS; \citealt{2001A&A...370L..49M}); 2)\, based on the centroid velocity ($\rm V_{lsr}$) of each field, we shifted the model spectrum and expanded it to a width of 50\,$\rm km\,s^{-1}$, ensuring clean channels that are free of multiple velocity components and broad spectral line wing emission.
The above approach yields line-free channels automatically.
Line emission channels were then flagged for continuum and line-free channels for four SPWs that were subtracted in the Fourier space employing the CASA task {\it uvcontsub} with a linear fitting (fitorder=1). 
We refer to this process as the {\it Auto-Flag} step. The combined imaging processes of both continuum and lines were performed using the CASA task {\it tclean}, with a {\it briggs} robust weighting of 0.5. 
This constitutes the first version of the TM2+ACA data for both continuum and spectral lines (see Figure\,\ref{fig:flowchat}).

However, the TM2+ACA first version data reveal a skewed baseline for some fields instead of a constant level. This could be attributed to complex molecular spectral lines or more extended line wings, which resulted in contaminated line-free channels generated in the {\it Auto-Flag} step.
Therefore, we manually reidentified the line-free channels for each field based on the TM2+ACA first version data using  {\it Linemarker}\,\footnote{\url{https://gitee.com/liuxunchuan/linemarker}; A tool for manually identifying the line-free or line-emission channels in an astronomical spectrum.}.
Line-free channels were then flagged before the next continuum and line imaging.  
This is refered to as the {\it Manual-Flag} step (see Figure\,\ref{fig:flowchat}).

In the image cleaning process, the image size for each field in Group\,1 was set to be [300, 300] pixels, with a pixel size of 0.15\arcsec; for the fields in Groups\,2-15 each image was set to have [250, 250] pixels, with a  pixel size of 0.2\arcsec.
The different settings mentioned above are due to the various angular resolutions between Group\,1 and other groups (see Table\,\ref{tab:almaobs}). The image cleaning procedure was conducted with in {\it pblimit} of 0.2 in each field.

In addition, as mentioned in Sec.\,\ref{subsec:quarks sample}, 17 protocluster clumps with extended density structures inside were covered by two single-pointings. For 15 of them, we made the mosaic before imaging. However, the I12572-6316 and I17269-3312 protocluster clumps were excluded for the mosaic imaging because of the insufficient overlap between the two associated single-pointings.

\begin{figure*}[ht!]
    \centering
    \includegraphics[angle=0, width=1.\textwidth]{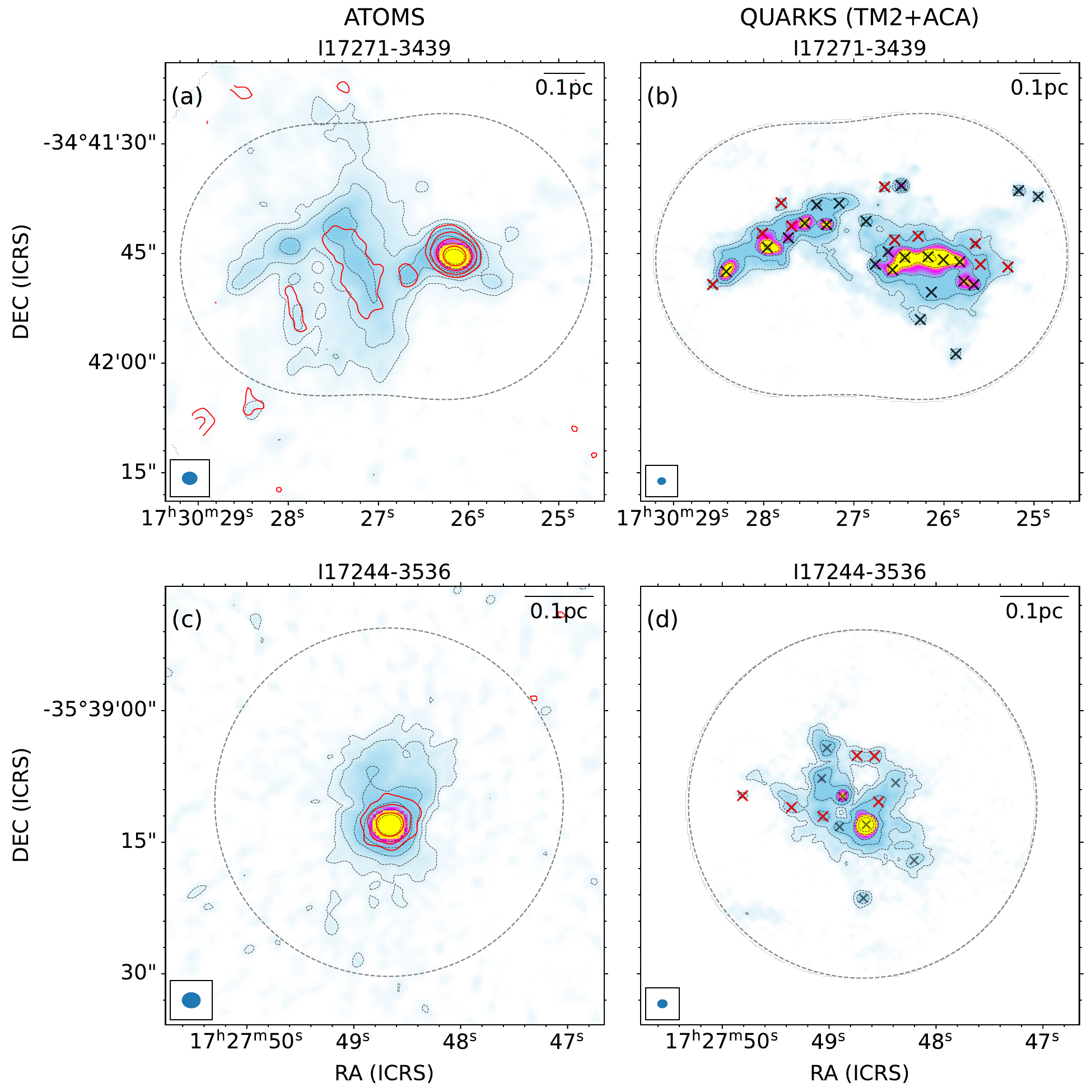}  
    \caption{Comparison of ATOMS 3\,mm and QUARKS 1.3\,mm continuum images for two protocluster clumps. The black dashed circle in each panel defines the field of view of the QUARKS survey. Left panels (a, c): ATOMS 3\,mm continuum emission. The dashed contour levels are [3, 6, 12, 24, 48, 96]\,rms, with $\rm rms\sim 1.4$ and $\rm \sim 0.4\,mJy\,beam^{-1}$ for I17271-3439 and I17244-3536, respectively. Red contours correspond to integrated emission of the $\rm H40\alpha$ recombination line representative of ionized gas from the ATOMS survey, with rms $\sim\rm 0.2$ and $\sim \rm 0.1\,Jy\,beam^{-1}\,km\,s^{-1}$ for I17271-3439 and I17244-3536, respectively.
    Right panels (b, d): QUARKS TM2+ACA 1.3\,mm continuum emission. The contour levels are the same as in left panels, but with $\rm rms \sim 2.1$ and $\sim \rm 0.4\,mJy\,beam^{-1}$ for I17271-3439 and I17244-3536, respectively.
    The cross symbols correspond to the cores extracted from 1.3mm continuum emission using the {\it getsf} algorithm, where black markers denote those detected by the algorithm at a configuration with a minimum source size employing the beam size, while red markers correspond to relatively faint cores detected at the other configuration employing half-beam.
    The beam sizes of ATOMS and QUARKS TM2+ACA continuum emission are shown on the lower left, and the 0.1\,pc scale bar is on the upper right of each panel.}
    \label{fig:mst}    
\end{figure*}

\subsubsection{Continuum imaging with self-calibration\label{sub:continuum}}
All four SPWs visibility data were used to ensure a rather high sensitivity for the QUARKS TM2+ACA continuum image. In addition, the self-calibration approach was applied to improve the dynamic range of the reduced image by correcting the visibility phase/amplitude via the comparison of the visibility data with the model of the source itself \citep{2022arXiv220705591R}.
For continuum imaging, we performed three rounds of phase self-calibration (see Fig.\,\ref{fig:flowchat}), each based on the {\it tclean} results of the previous round. Here, we selected the {\it hogbom} deconvolving algorithm which uses a point source model of the sky brightness distribution for operation, and is a better choice in the iterative phase self-calibration process. Subsequently, the Multi-Term Multi-Frequency Synthesis ({\it mtmfs}) with an {\it nterm} of 2 was employed for the last {\it tclean} procedure, which both restores the extended structure and improves the image quality \citep{2011A&A...532A..71R}.
Self-calibration was carried out manually and iteratively to optimize the cleaning thresholds, ensuring a rather good sensitivity while preventing from the divergence.
As a result, the average noise level of the reduced TM2+ACA continuum data is $\sim\rm 0.6\,mJy\,beam^{-1}$, with an average beam size of $\sim1.3$\,\arcsec$\times1.1$\,\arcsec, corresponding to a mass sensitivity of $\sim 0.3$\,\msun\ for a 30\,K dust temperature at a median distance of 3.7\,kpc for the QUARKS survey.

\subsubsection{Line Imaging}
For molecular lines, we applied to four SPWs all the calibration tables from three rounds of continuum self-calibration (see Sect.\,\ref{sub:continuum}), and subtracted  emission of line-free channels in the Fourier space by employing the CASA task {\it uvcontsub} with a linear fitting (fitorder=1) (see Fig.\,\ref{fig:flowchat}).
The subtracted TM2 and ACA visibility data were concatenated and subsequently cleaned to generate the spectral cube using the CASA task {\it tclean}, with the {\it mutilscale} deconvolution approach and a uniform cleaning threshold of 25$\,\rm mJy\,beam^{-1}$ ($\sim \rm 0.4\,K$). This threshold was chosen to optimize the cleaning performance of spectral cubes for most fields. As a result, the average rms level of the reduced line data is $\sim \rm 10\,mJy\,beam^{-1}$ ($\sim \rm 0.2\,K$) per channel at an average synthesized beam of $\sim 1.4$\,\arcsec$\times1.1$\,\arcsec\ and a velocity resolution of $\sim\rm 1.3\,km\,s^{-1}$.

\begin{deluxetable*}{cccccccccccc}\label{table:getsf}
\tabletypesize{\scriptsize}
% \tablewidth{0pt} 
\tablecaption{Core parameters measured by {\it getsf}.} 
% \tablecomments{}
% \centering
\tablehead{
\colhead{Source Name} & \colhead{Core} & \colhead{Dist.}    & \colhead{RA\,(IRCS)} & \colhead{Dec\,(IRCS)} & \colhead{Maj} & \colhead{Min} & \colhead{PA} & \colhead{$\rm F^{int\,\it(a)}_{1.3mm}$} & \colhead{$\rm F_{1.3mm}^{peak\,\it(a)}$}& \colhead{$SIGNM$}& \colhead{$GOODM$}\\
&& \colhead{(kpc)}& \colhead{(h:m:s)}  & \colhead{(d:m:s)} & \colhead{($\arcsec$)} & \colhead{($\arcsec$)} & \colhead{($\degree$)} & \colhead{(mJy)} & \colhead{$(\rm mJy\,beam^{-1})$}  \\ }
\colnumbers 
\startdata 
I08303-4303 & 1 & 2.35 & 8h32m08.64s & -43d13m45.65s & 1.195 & 0.9643 & 27.43 & 134.16(1.48) & 14.59(0.51) & 171 & 705  \\
I08303-4303 & 2 & 2.35 & 8h32m09.05s & -43d13m43.33s & 0.9634 & 0.8254 & 93.93 & 33.38(1.08) & 5.96(0.56) & 74 & 116   \\
I08303-4303 & 3 & 2.35 & 8h32m08.46s & -43d13m49.13s & 1.453 & 1.039 & 38.52 & 52.66(1.4) & 5.88(0.6) & 59 & 82   \\
I08303-4303 & 4 & 2.35 & 8h32m08.46s & -43d13m47.22s & 1.081 & 1.029 & 0.6709 & 28.22(1.06) & 4.17(0.62) & 31 & 40  \\
I08303-4303 & 5 & 2.35 & 8h32m08.90s & -43d13m52.12s & 1.148 & 0.8792 & 11.07 & 6.17(0.38) & 1.03(0.19) & 12 & 8   \\
I08303-4303 & 6 & 2.35 & 8h32m08.88s & -43d13m42.05s & 2.086 & 1.466 & 99.41 & 7.55(0.91) & 0.48(0.37) & 13 & 3  \\
\enddata
\begin{flushleft}
(The complete table is available in machine-readable form.)\\
$^{(a)}$ The values in parentheses represent the measurement errors.
\end{flushleft}

\end{deluxetable*}

% \begin{deluxetable*}{cccccccccc}\label{table:getsf}
% % \tabletypesize{\scriptsize}
% % \tablewidth{0pt} 
% \tablecaption{Core parameters measured by {\it getsf}.} 
% % \tablecomments{}
% % \centering
% \tablehead{
% \colhead{Source Name} & \colhead{Core} & \colhead{Dist.}    & \colhead{RA\,(IRCS)} & \colhead{Dec\,(IRCS)} & \colhead{Maj} & \colhead{Min} & \colhead{PA} & \colhead{$\rm F^{int\,\it(a)}_{1.3mm}$} & \colhead{$\rm F_{1.3mm}^{peak\,\it(a)}$}\\
% && \colhead{(kpc)}& \colhead{(h:m:s)}  & \colhead{(d:m:s)} & \colhead{($\arcsec$)} & \colhead{($\arcsec$)} & \colhead{($\degree$)} & \colhead{(mJy)} & \colhead{$(\rm mJy\,beam^{-1})$}  \\ }
% \colnumbers 
% \startdata 
% I08303-4303 & 1 & 2.35 & 8h32m08.64s & -43d13m45.65s & 1.2 & 1.0 & 27.4 & 134.2\,(1.5) & 14.6\,(0.5) \\
% I08303-4303 & 2 & 2.35 & 8h32m09.05s & -43d13m43.33s & 1.0 & 0.8 & 93.9 & 33.4\,(1.1) & 6.0\,(0.6) \\
% I08303-4303 & 3 & 2.35 & 8h32m08.46s & -43d13m49.13s & 1.5 & 1.0 & 38.5 & 52.7\,(1.4) & 5.9\,(0.6) \\
% I08303-4303 & 4 & 2.35 & 8h32m08.46s & -43d13m47.22s & 1.1 & 1.0 & 0.7 & 28.2\,(1.1) & 4.2\,(0.6) \\
% I08303-4303 & 5 & 2.35 & 8h32m08.90s & -43d13m52.12s & 1.1 & 0.9 & 11.1 & 6.2\,(0.4) & 1.0\,(0.2) \\
% I08303-4303 & 6 & 2.35 & 8h32m08.88s & -43d13m42.05s & 2.1 & 1.7 & 99.4 & 7.6\,(0.9) & 0.5\,(0.4) \\
% \enddata
% \begin{flushleft}
% (The complete table is available in machine-readable form.)\\
% $^{(a)}$ The values in parentheses represent the measurement errors.
% \end{flushleft}

% \end{deluxetable*}

\section{Results and analysis}\label{sec:result}
\subsection{Core extraction}\label{sub:sec:res:core_extraction}
Figure\,\ref{fig:mst} presents a comparison of continuum images between the 3\,mm ATOMS data (left panels) and the 1.3\,mm QUARKS TM2+ACA data (right panels) for two example targets. The figure demonstrates the overall spatial consistency between 1.3\,mm and 3\,mm emission. The QUARKS images, owing to their higher angular resolution, reveal finer structures such as dense filaments and cores. In contrast, ATOMS 3\,mm emission appears more extended, which results from the larger maximum recoverable scale of the ATOMS survey ($\sim 20$\,\arcsec) and/or contribution from free-free emission at 3\,mm.
Free-free emission from ionized gas could radiate significantly at 3\,mm particularly within rather evolved clumps associated with UCH{\sc ii} regions, while at 1.3\,mm, thermal dust emission could be the major component (e.g., \citealt{2020ApJS..248...24G,2022A&A...662A...8M}).
 As shown in the right panels of Figure\,\ref{fig:mst}, the red contours correspond to  emission of the $\rm H40\alpha$ recombination line representative of ionized gas from the ATOMS survey, which does show a somehow extended emission morphology.

We employed the {\it getsf} algorithm \citep{2021A&A...649A..89M} to identify compact structures within the TM2+ACA continuum data at 1.3\,mm. This algorithm has been widely used for extracting compact sources in complex protocluster environments \citep{2022A&A...664A..26P,2023A&A...674A..75N,2024MNRAS.527.5895D,2024ApJS..270....9X,2024A&A...690A..33L,2024ApJ...976..241Y,xu2025duet}. In our case, the {\it getsf} was applied to images without primary beam correction, as they exhibit a relatively uniform noise level across the entire field of view. For the algorithm's configuration, we set the minimum and maximum sizes of the extracted compact sources to half and three times the synthesized beam size, respectively. Note that the choice of minimum size (whether half- or one-beam size) does not significantly influence the parameters (e.g., size, flux) of the extracted sources with strong emission, but the former parameter configuration can extract relatively weaker structures, which enables a sample of fragments as complete as possible for clump-to-core fragmentation analysis. As shown in the right panels of Figure\,\ref{fig:mst}, the red cross symbols represent the additional faint structures identified by the {\it getsf} using half of the beam size as the minimum size.

To enhance the reliability of our core sample, we applied a post-selection process based on the quality-check parameters provided by the {\it getsf} algorithm. Specifically, we retained only the cores with a significance level ({\it SIGNM}; detection significance from monochromatic single scales) greater than 5 and a goodness ({\it GOODM}; monochromatic goodness combining significance and signal-to-noise ratio) greater than unity \citep{2018A&A...619A..52B}, while excluding the cores located at the edges of each field. 
Cores with a value of {\it GOODM} below unity were considered highly unreliable \citep{2021A&A...649A..89M}. 
The {\it SIGNM} and {\it GOODM} values for each core are listed in Col.\,11-12 of Table\,\ref{table:getsf}.

Following the above criteria, a total of 1562 compact structures (i.e., cores) were identified across 156 fields. The core parameters derived from {\it getsf}, with the integrated flux and peak intensity corrected for the primary beam response, are presented in Table\,\ref{table:getsf}. It is worth noting that although the majority of compact cores in each field can be identified by {\it getsf} using our criteria, a few potential sources could be still missing. As shown in Figure\,\ref{fig:mst}d for an example, two or three relatively weak continuum peaks lack corresponding compact core identifications. This arises due to their low intensity contrast relative to background emission, and therefore those weak density structures could not be recognized by {\it getsf}  \citep{2021A&A...649A..89M}. Except for I13134-6242 and I18056-1952 (each with 3 dense cores), our clump sample contains 4 to 33 cores, which provides rather sufficient core numbers for our following fragmentation analysis.

\begin{table*}[!thb]
\centering
\caption{Criteria of Core Classification. \label{tab:classification:criterion}}
\setcellgapes{1.2pt}
% \makegapedcells
% \linespread{1.2}
\renewcommand{\arraystretch}{1.8} % Default value: 1
\begin{tabular}{lll}
\hline
\hline
 Type of Cores & Criteria & Assumed temperature \\
% (1) & (2)\\
\hline
Candidate Starless Core (SC) & $N_{\rm line}\le\,6$ without outflows and $\rm H30\alpha$ emission & [20\,K, \,$T_{\rm clump}$]$^{a}$\\
\hline
Candidate Warm Core (WC) & $\rm 6< N_{line}\le\,15$ without $\rm H30\alpha$ emission&30\,K\\
& Or $N_{\rm line}\le\,6$ with outflows but no $\rm H30\alpha$ emission\\ 
\hline
Evolved Cores & &100\,K\\
Candidate Hot Molecular Core (HMC) & $N_{\rm line}\,>\,15$ without $\rm H30\alpha$ emission\\
Candidate UCH{\sc ii}-{\it h} Core  &  with $\rm H30\alpha$ emission and high line richness ($N_{\rm line}\,>\,15$)\\
Candidate UCH{\sc ii}-{\it l} Core & with $\rm H30\alpha$ emission and low line richness ($N_{\rm line}\le\,15$ ) \\
\hline
\end{tabular}{
\begin{flushleft}
$^a$\,$T_{\rm clump}$ is the average dust temperature of the natal clump.
\end{flushleft}
}
\end{table*}

\subsection{Classification of cores}\label{sub:sec:res:classification}
We used molecular line emission from four SPWs in the QUARKS survey to classify the cores into distinct evolutionary stages of star formation. The four SPWs cover a wide frequency range of approximately 8\,GHz, encompassing numerous molecular transitions that probe star-forming environments across various evolutionary stages, from starless cores to UCH{\sc ii} regions.
The classification criteria are summarized in Table\,\ref{tab:classification:criterion}, with further details on each category of cores provided below.
The number of detectable molecular lines ($N_{\rm line}$) serves as a quantitative and efficient metric to assess the chemical complexity of the cores, enabling an approximate determination of their evolutionary stage \citep{2021MNRAS.505.2801L}.

To determine $N_{\rm line}$ for each core, we identified the number of line emission peaks from the core-averaged spectra exceeding a $3\sigma$ noise level across the four SPWs. The $1\sigma$ noise level was calculated as the standard deviation of the amplitudes in the line-free emission channels for each core ($\sim\rm 20\,mJy\,beam^{-1}$ on average). The resulting $N_{\rm line}$ values for each core are listed in Column\,6 of Table\,\ref{table:phypre} and Table\,\ref{table:phywarm}.
Note that $N_{\rm line}$ could be underestimated for cores having a large number of complex molecular transitions (e.g., $N_{\rm line}$>50), though this does not affect our major classification results.

Using $N_{\rm line}$ in conjunction with other star formation indicators (e.g., outflows, and ionized gas), we categorized the 1562 identified cores into three primary groups: 1) candidate starless cores, 2) candidate warm cores (WC), and 3) candidate evolved cores. The evolved core category was further subdivided into three subcategories based on $\rm H30\alpha$ ionized gas emission: candidate hot molecular cores (HMCs), candidate UCH{\sc ii} region cores with a relatively high line richness ($N_{\rm line} > 15$; UCH{\sc ii}-{\it h}), and candidate UCH{\sc ii} region cores with a relatively low line richness ($N_{\rm line} \leq 15$; UCH{\sc ii}-{\it l}). Assuming that stronger ionized gas feedback occurring in later  evolutionary stages reduces the richness of molecular line emission (e.g., \citealt{2006A&A...455..971R,2008ApJ...675L..33B,2015ApJ...803...39Q}), UCH{\sc ii}-{\it l} cores could be considered more evolved than UCH{\sc ii}-{\it h} ones.

\begin{figure}[ht!]
    \centering
    \includegraphics[angle=0, width=0.45\textwidth]{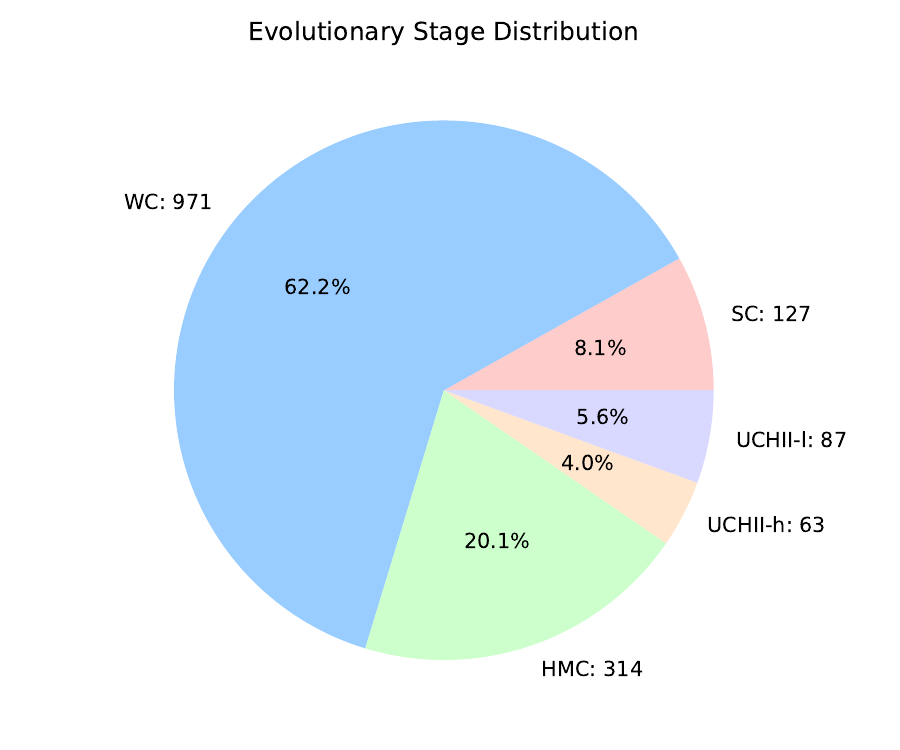} 
    \caption{Pie chart for statistics of core classifications.} 
    \label{fig:piechart}    
\end{figure}

\begin{figure*}[ht!]
    \centering
    \includegraphics[angle=0, width=0.93\textwidth]{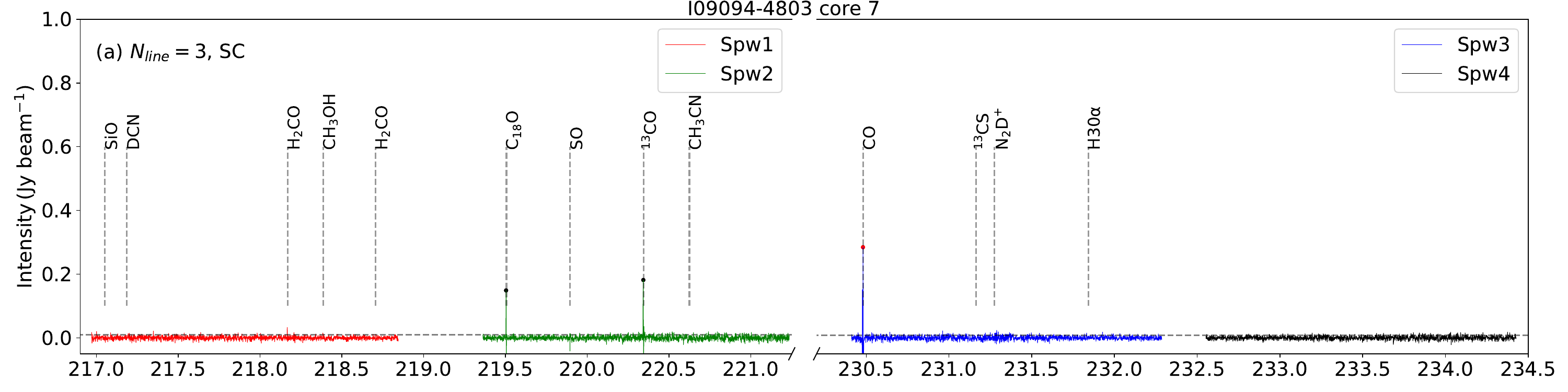} 
    \includegraphics[angle=0, width=0.93\textwidth]{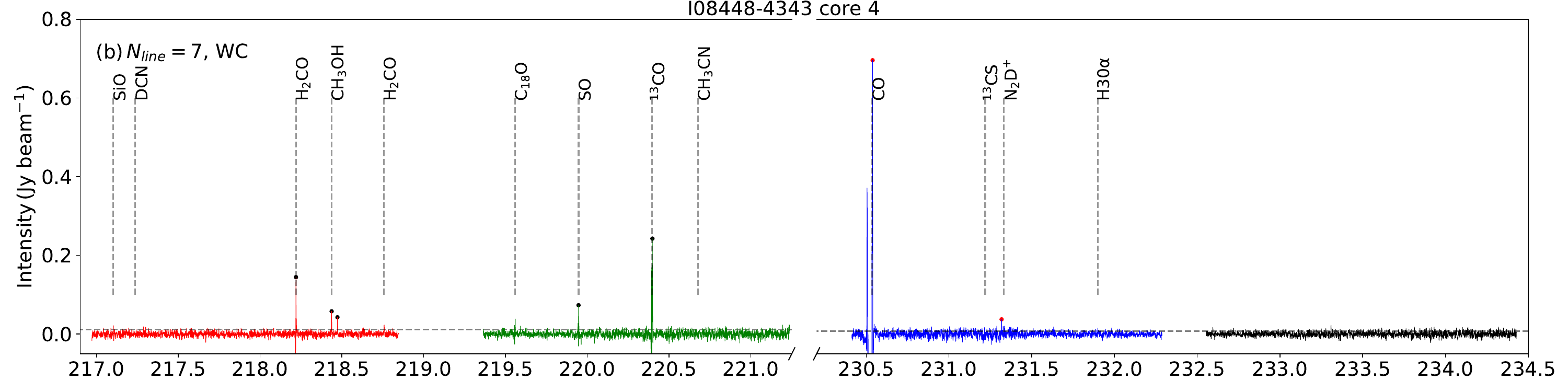} 
    \includegraphics[angle=0, width=0.93\textwidth]{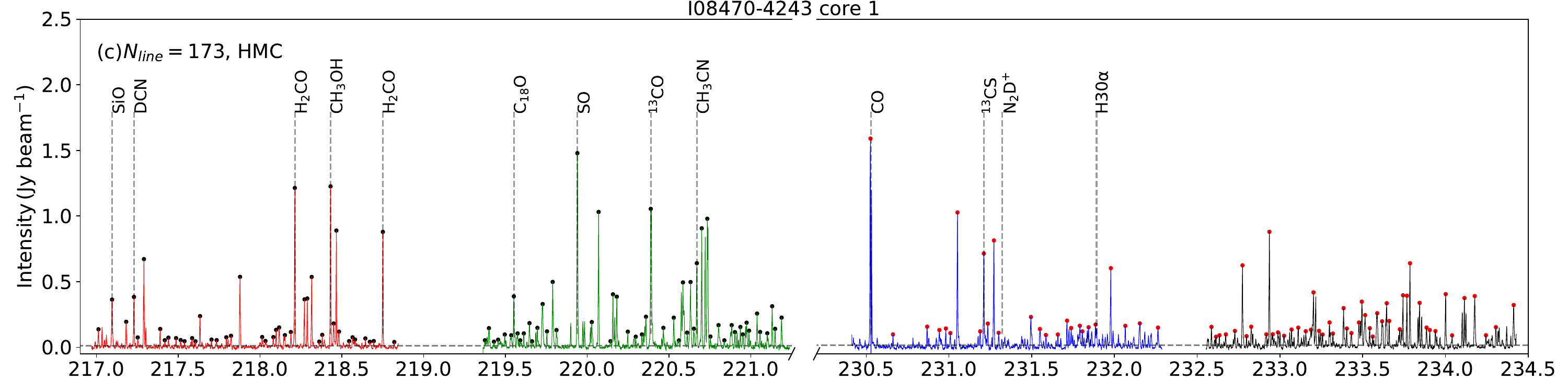} 
    \includegraphics[angle=0, width=0.93\textwidth]{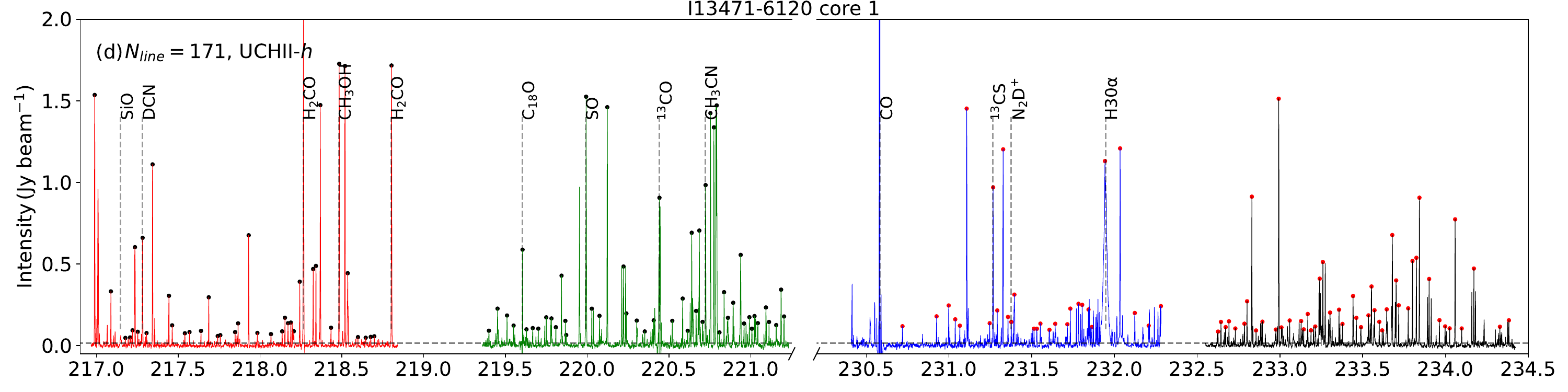}
    \includegraphics[angle=0, width=0.93\textwidth]{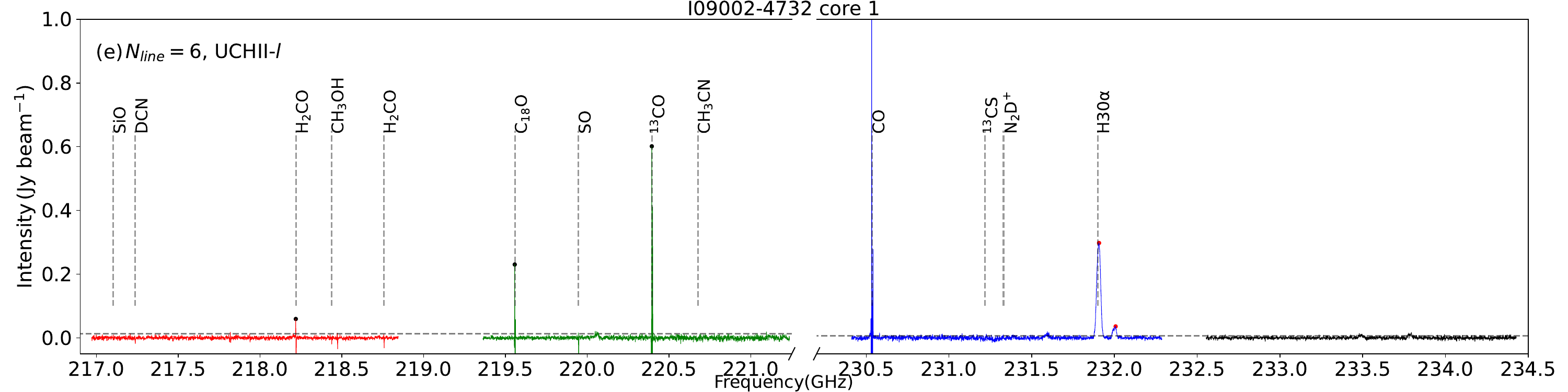}
    \caption{Demonstration of core-averaged spectra covered in the four spectral windows of the QUARKS survey for dense cores representative of different stages of evolution, including candidates of (a) starless core, (b) warm core, (c) hot core, (d) UCH{\sc ii}-{\it h} core and (e) UCH{\sc ii}-{\it l} core. On the upper left corner of each panel displays the count of molecular spectral lines tentatively detected in each core. The horizontal dashed line represents the noise level of the spectra. Several common molecular line transitions are labeled. } 
    \label{fig:fourspws}    
\end{figure*}

\begin{figure*}[ht!]
    \centering
    \includegraphics[angle=0, width=0.4\textwidth]{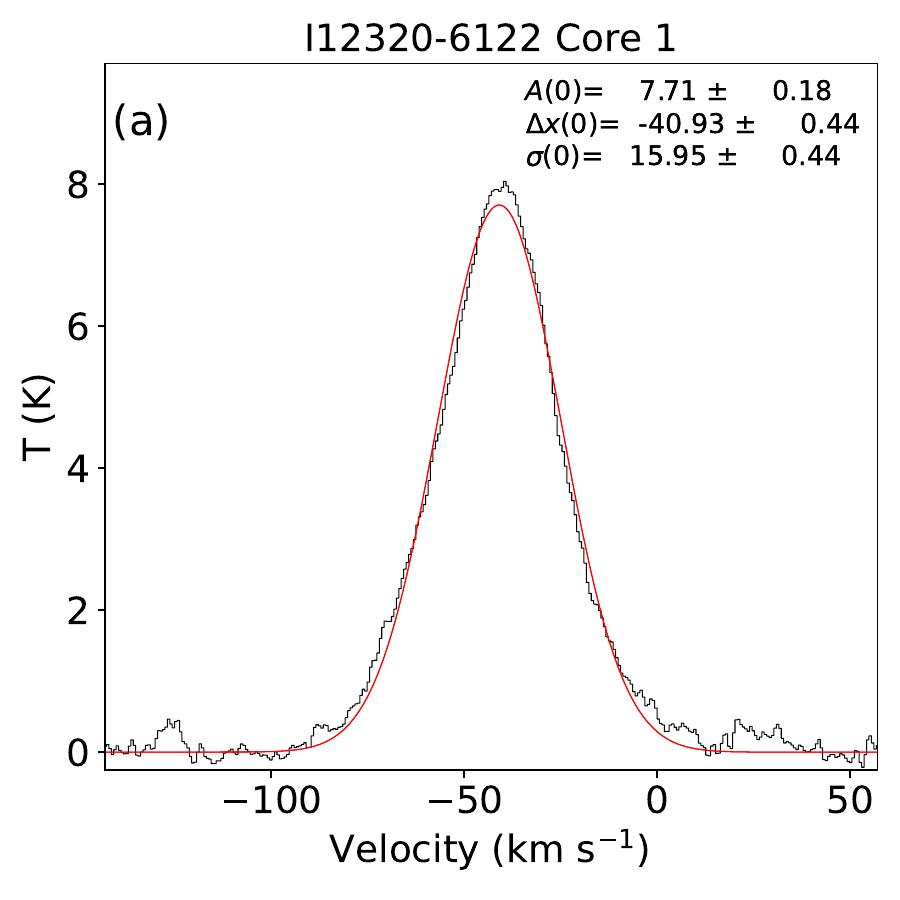} 
    \includegraphics[angle=0, width=0.45\textwidth]{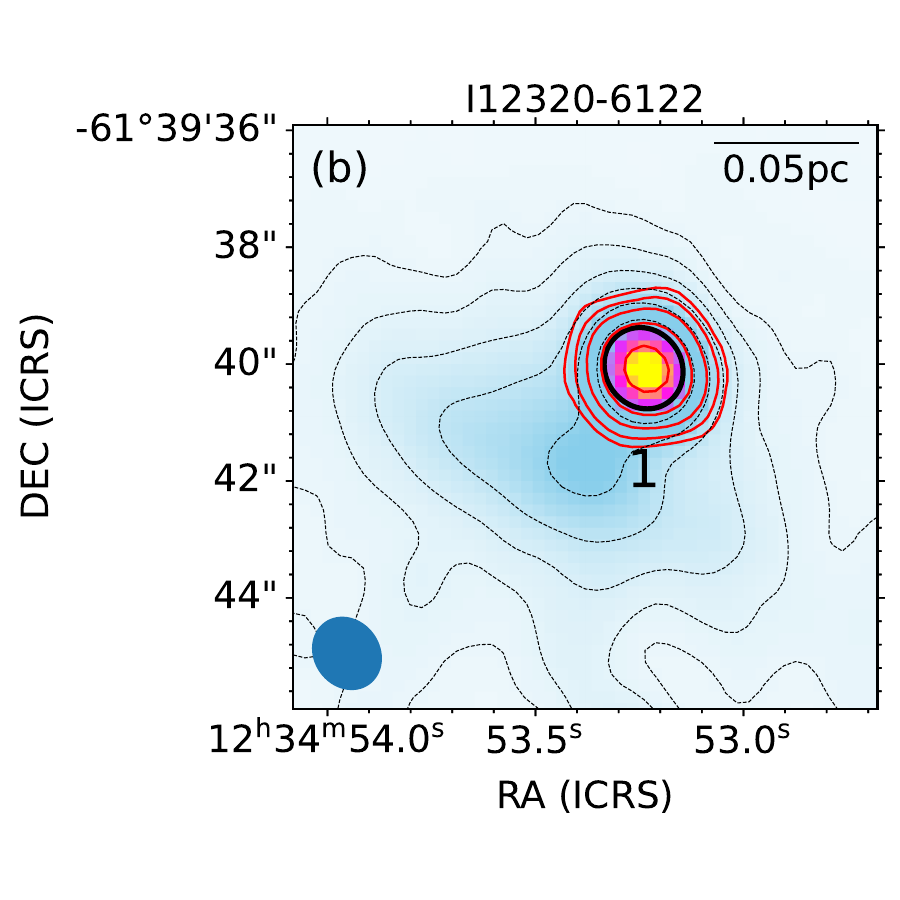} 
    \caption{Emission of the $\rm H30\alpha$ line toward I12320-6122 Core\,1. Panel\,(a): The black histogram represents the observed core-averaged spectrum  while the red line does the Gaussian fitting. Panel\,(b): QUARKS TM2+ACA continuum image of I12320-6122 Core\,1. The contour levels are the same as in Figure\,\ref{fig:mst}, but with $\rm rms=1\,mJy\,beam^{-1}$. The red contours show velocity-integrated emission of $\rm H30\alpha$ over [-90, 10]\,$\rm km\,s^{-1}$, with $\rm rms=0.5\,Jy\,beam^{-1}\,km\,s^{-1}$. The black circle delineates I12320-6122 Core\,1. The beam size of the TM2+ACA observations is shown on the lower left, and the 0.05\,pc scale bar is on the upper right corner. }
    \label{fig:h30}    
\end{figure*}

\subsubsection{Candidate starless and warm cores}\label{sub:sub:sec:pwc}
Candidate starless cores are defined here based on the two following criteria. First, they exhibit gas emission from some molecular transitions covered in the QUARKS survey with upper energy levels ($E_u$) below 22\,K \citep{2019ApJ...886..102S}. In accordance with this, six line transitions are considered relevant, namely CO/$^{13}$CO/C$^{18}$O~(2-1) with $E_u\sim 16\rm\,K$, H$_2$CO~(3$_{0,3}$–2$_{0,2}$) with $E_u=20.96\rm\,K$, DCN (3–2) with  $E_u=20.85\rm\,K$, and N$_2$D$^+$ (3–2) with $E_u=22.2\rm\,K$. Thus, candidate starless cores are required to only have such cold gas emission of lines with $N_{\rm line} \leq 6$. Second, such candidates should lack any star formation signatures, such as outflows traced by CO emission or ionized gas by H30$\alpha$ emission (e.g., \citealt{2018ApJS..235....3Y,2019ApJ...886..102S,2022A&A...658A.160Y,2022MNRAS.510.3389U}). 
Following the above standard, 127 (8.1\%) out of 1562 cores were finally classified as candidate starless cores (see Figure\,\ref{fig:piechart}).
In addition, H$_2$CO~(3$_{0,3}$–2$_{0,2})$ and  DCN (3–2) line emission could  trace low-velocity outflows \citep{2012ApJ...757L...9C}. However, by examining the core-averaged spectra of these two lines for the 127 starless candidates, no broad line-wings were observed in the spectral line profiles (not shown here). This further suggests  the starless possibility of these cores.
Figure\,\ref{fig:fourspws}a presents typical molecular line spectra for a candidate starless core, which shows a few lines not signifying any star-forming signatures as mentioned earlier.

Warm cores are defined here to represent a transitional phase between starless and hot molecular cores. In this study, molecular transitions with upper energy levels below $\sim100$\,K typical of hot gas tracers but above $\sim 22$\,K typical of cold gas tracers, such as CH$_3$OH (4–3) with  $E_u=45.46\rm\,K$, H$_2$CO (3$_{2,2}$–2$_{2,1}$/3$_{2,1}$–2$_{2,0}$) with $E_u=68.09\rm\,K$, SiO (5–4) with $E_u=31.26\rm\,K$, SO (6–5) with $E_u=34.98\rm\,K$, and $^{13}$CS (5–4) with $E_u=33.29\rm\,K$, are approximately designated as warm gas tracers (e.g., \citealt{2019ApJ...886..102S}). This rough definition allows us to estimate the number of emission lines for candidate warm cores to be  $6 < N_{\rm line} \leq 15$.
Note that when $N_{\rm line}>15$, the molecular lines already contain hot gas tracers (e.g., $\rm CH_3CN$).
Additionally, those with emission of lines of $N_{\rm line} \leq 6$ associated with CO outflows but no $\rm H30\alpha$ emission are also considered candidate warm cores. As a result, 971 cores (62.2\%) were identified as candidate warm cores (see Figure\,\ref{fig:piechart}). Figure\,\ref{fig:fourspws}\,b shows representative molecular line spectra for a candidate warm core.

\subsubsection{Candidate hot molecular and UCH{\sc ii} cores}

In addition to the candidate starless and warm cores described above, the remaining cores are assumed to be in a more advanced evolutionary stage, due to their association with detection of rich molecular line transitions ($N_{\rm line} > 15$) and/or the $\rm H30\alpha$ line representative of ionized gas emission. 
A broad spectrum profile of the $\rm H30\alpha$ line (e.g., line width $\rm >15\,km\,s^{-1}$, e.g., \citealt{2021MNRAS.505.2801L}) and  a compact emission pattern in the line intensity map are both required to diagnose the signature of UCH{\sc ii} regions, as illustrated in Figure\,\ref{fig:h30}. Consequently, the remaining 464 cores ($\sim$30\%) were classified as evolved cores (see Figure\,\ref{fig:piechart}).
As previously mentioned, these evolved cores were further subdivided into three subcategories based on their associated $N_{\rm line}$ and detection of $\rm H30\alpha$ (see Table\,\ref{tab:classification:criterion}): 314 candidate HMCs without associated $\rm H30\alpha$ emission; 63 UCH{\sc ii}-{\it h} candidates with associated $\rm H30\alpha$ emission and $N_{\rm line} > 15$; and 87 UCH{\sc ii}-{\it l} candidates with associated $\rm H30\alpha$ emission and $N_{\rm line} \leq 15$ (see Figure\,\ref{fig:piechart}).
In the ATOMS survey, \cite{2021MNRAS.505.2801L} used $\rm H{40\alpha}$ emission to identify UCH{\sc ii} cores, detecting 89 sources. They also have associated $\rm H30\alpha$ emission, except for two sources within I16158-5055 and I16506-4512 protocluster clumps (see Table\,1 of \citealt{2021MNRAS.505.2801L}) that were not covered by the QUARKS observations. The typical molecular transitions of the evolved cores are illustrated in Figure\,\ref{fig:fourspws}c–e.

\subsection{Physical parameters of dense cores}\label{sub:sec:res:physicalpara}
The dust temperature ($T_{\rm dust}$) of a core is a fundamental parameter which subsequently determines other key properties of the core such as the mass and the density. However, accurately determining $T_{\rm dust}$ for individual cores within a clump remains challenging, particularly for candidate starless and warm cores lacking detection of temperature-sensitive line transitions (e.g., CH$_3$CN K-ladder transitions). Besides, assigning a uniform $T_{\rm dust}$ to all cores within a clump is unrealistic due to their probable diverse evolutionary stages (see Sect.\,\ref{sub:sec:res:classification}). Instead, we assigned $T_{\rm dust}$ for cores based on their evolutionary stage, as shown in Column\,3 of Table\,\ref{tab:classification:criterion}.

For starless cores without internal protostellar heating, dust temperatures as low as 10–15\,K have been observed in IRDC environments \citep{2013ApJ...773..123S,2023A&A...675A..53B,2023ApJ...950..148M,2024ApJ...961L..35M}. However, given the IR-bright nature of the QUARKS sample \citep{2024RAA....24f5011X,2021MNRAS.505.2801L}, these temperatures could be higher (e.g., $\sim20$\,K; \citealt{2018A&A...618L...5N,2023A&A...674A..75N}) due to external heating by nearby young stars \citep{2024A&A...687A.217D}. Therefore, we adopted here a lower limit of temperature of 20\,K for all candidate starless cores, consistent not only with the assumption made in the study by the ALMA-IMF survey team \citep{2025arXiv250209426V} but also with the averaged temperature measured in starless cores within clustered star-forming regions (Table\,5 of \citealt{2013MNRAS.432.3288S}). For the upper limit of temperature, we assigned the average $T_{\rm dust}$ of the natal clump, derived from the spectral energy distribution (\citealt{2004A&A...426...97F}; see Table\,A1 in \citealt{2020MNRAS.496.2790L}).

For candidate warm cores, a dust temperature of 30\,K was adopted \citep{2023MNRAS.520.2306T}, consistent with the average $T_{\rm dust}$ ($\sim28.6$\,K) of their natal clumps. Finally, for candidate evolved cores, we assumed a dust temperature of 100\,K \citep{2009ARA&A..47..427H,2022MNRAS.511.3463Q,2023ApJ...950...57T}.

Assuming dust continuum emission at 1.3\,mm  to be optically thin, the masses of the cores can be estimated as follows
\begin{equation}\label{equ:mass}
    M_{\rm core}=\frac{S^{\rm int}_{\nu}D^2R_{\rm gd}}{\kappa_vB_v(T_{\rm dust})},
\end{equation}
where the integrated flux ($S_{\nu}^{\rm int}$) is measured from 1.3\,mm dust emission, $D$ is the distance for each source, adopted from \cite{2024RAA....24b5009L}. $R_{\rm gd}$ is the gas-to-dust ratio, assumed to be 100 (e.g., \citealt{2019ApJ...886..102S,2022A&A...662A...8M,2023MNRAS.520.2306T,2025arXiv250305663C}) \footnote{\label{foot2} Other values for $R_{\rm gd}$ can be found in literature, ranging from 110--150 (e.g., \citealt{2018A&A...617A.100B,2019ApJ...886...36S,2021MNRAS.503.4601B,2023A&A...675A..53B}). This implies an $R_{\rm gd}$ uncertainty of $\sim50\%$.}. The dust opacity ($\kappa_{\nu}$) is taken as 0.9\,$\rm cm^{2}\,g^{-1}$ \citep{1994A&A...291..943O} and $B_{\rm \nu}(T_{\rm dust})$ is the Planck function for a given $T_{\rm dust}$.

\begin{figure*}[ht!]
    \centering
    \includegraphics[angle=0, width=1\textwidth]{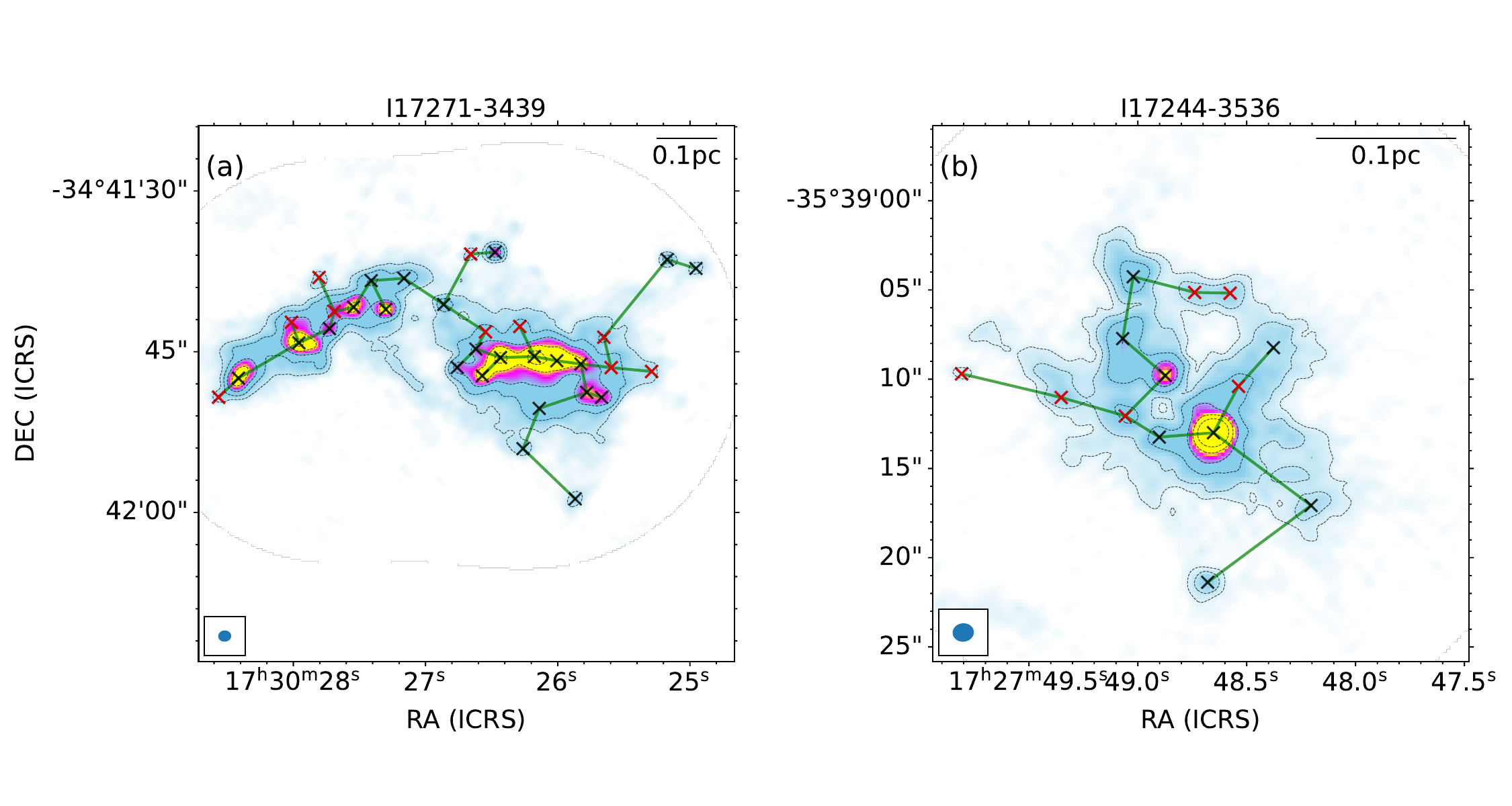} 
    \caption{Example of the MST results (green segments).  QUARKS TM2+ACA 1.3mm continuum emission is shown in colorscale with dense cores (i.e., cross symbols) overlaid. The contour levels are the same as in Figure\,\ref{fig:mst}b,\,d. 
     The beam size of the QUARKS data is shown on the lower left, and the 0.1\,pc scale bar is on the upper right.} 
    \label{fig:MST_example}    
\end{figure*}

Assuming a spherical geometry, the number density of the cores was calculated as $n_{\rm H_2} = \frac{3M_{\rm core}}{4\pi R_{\rm c}^3 \mu m_{\rm H}}$, where $\mu = 2.8$ is the mean molecular weight and $m_{\rm H}$ is the hydrogen atom mass \citep{2008A&A...487..993K}. The surface density is given by $\Sigma = \frac{M_{\rm core}}{\pi R_c^2}$. The deconvolved effective radius ($R_c$) for each core was determined using $R_c = \sqrt{(maj \times min) - (bmaj \times bmin)}/2 \times D$ \citep{2022A&A...664A..26P}, where $maj$ and $min$ are the sizes of the major and minor axes measured by the {\it getsf} algorithm (see Columns 6–7 in Table\,\ref{table:getsf}), and $bmaj$ and $bmin$ correspond to the synthesized beam dimensions of the QUARKS TM2+ACA observations. To mitigate deconvolution artifacts, a minimum radius of half the beam size ($\sim0.5$\,\arcsec) was enforced, following \citet{2022A&A...664A..26P}.

Considering the uncertainties in distance ($\rm\sim 10\%$), dust temperature ($\rm <30\%$) alongwith that from $\kappa_\nu$ and $R_{\rm gd}$ ($\sim 50\%$; \citealt{2017ApJ...841...97S,2024ApJ...974..239S}), we estimated the uncertainty of mass, number density, and surface density of $\rm\sim 50\%$, using the Monte Carlo approach as adopted in \citet{2024ApJ...974..239S}.

As shown in Table\,\ref{table:phypre}, candidate starless cores have masses (at 20\,K) ranging from [0.2–21.0]\,M$_{\odot}$, with a median of 1.5\,M$_{\odot}$, which are listed in Column\,8 of Table\,\ref{table:phypre}. Their estimated masses, calculated using the average dust temperature of their natal clumps, are listed in Column\,9 of Table\,\ref{table:phypre}. The median radius, number density, and surface density of candidate starless cores are 2800\,au, $2.0 \times 10^6$\,cm$^{-3}$, and 0.5\,g~cm$^{-2}$, respectively (based on a temperature of 20\,K).

The physical parameters for candidate warm and evolved cores are summarized in Table\,\ref{table:phywarm}. For candidate warm cores, the median radius, number density and surface density are 2.1\,\msun\,, 2800\,au, $2.6 \times 10^6$\,cm$^{-3}$, and 0.7\,g~cm$^{-2}$, respectively. For evolved cores, the median radius, number density and surface density are 3.7\,\msun\,, 2700\,au, $5.5 \times 10^6$\,cm$^{-3}$, and 1.4\,g~cm$^{-2}$.
Figure\,\ref{fig:corephysical} illustrates the distribution of physical parameters across different categories of cores studied here. Although there is a clear increase in the total flux of cores as they evolve (see Figure\,\ref{fig:corephysical}\,a), no significant differences are observed in other parameters (i.e., mass, radius, number density, and surface density, see Figure\,\ref{fig:corephysical}\,b-e) amongst cores in different evolutionary stages.

\begin{figure}[ht!]
    \centering
    \includegraphics[angle=0, width=0.48\textwidth]{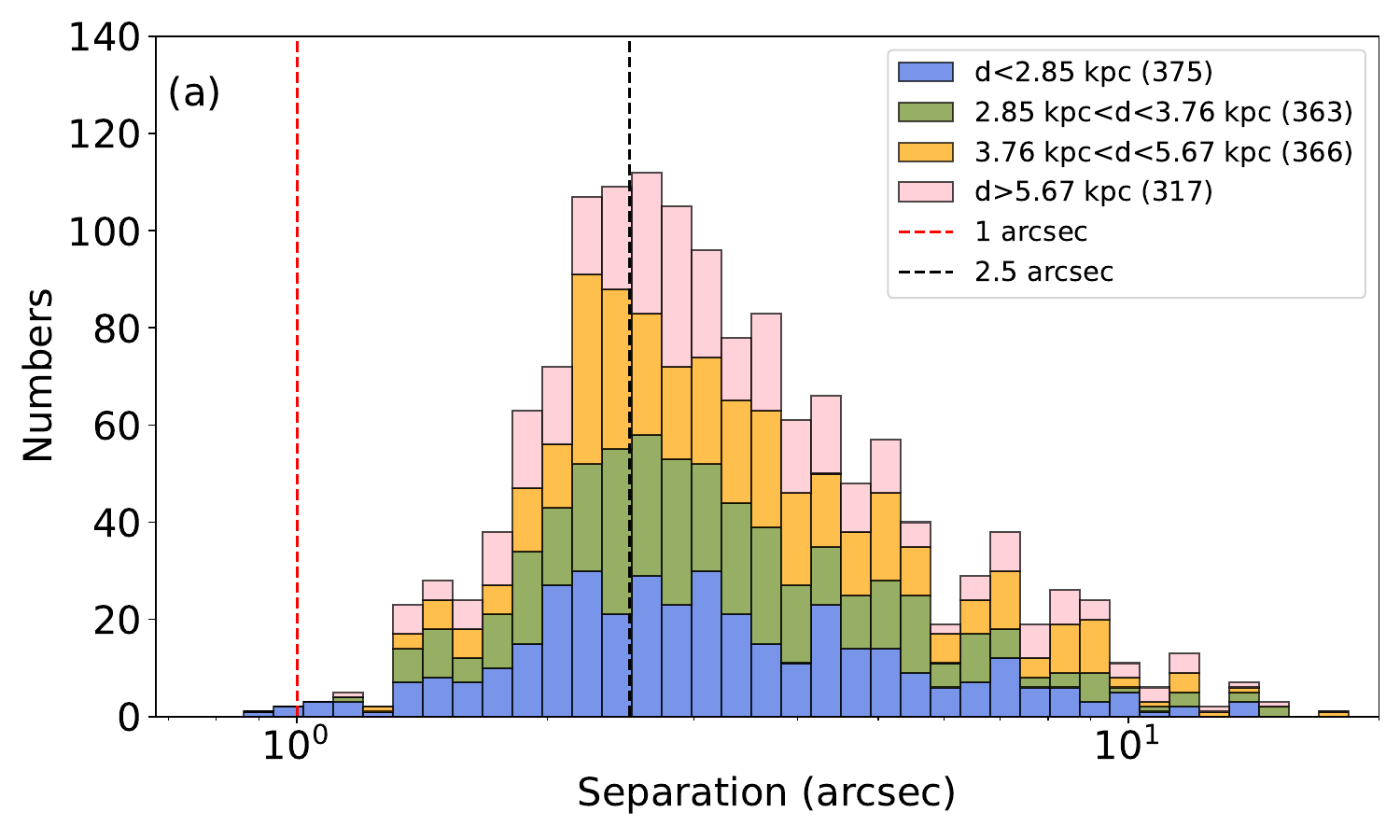}
    \includegraphics[angle=0, width=0.48\textwidth]{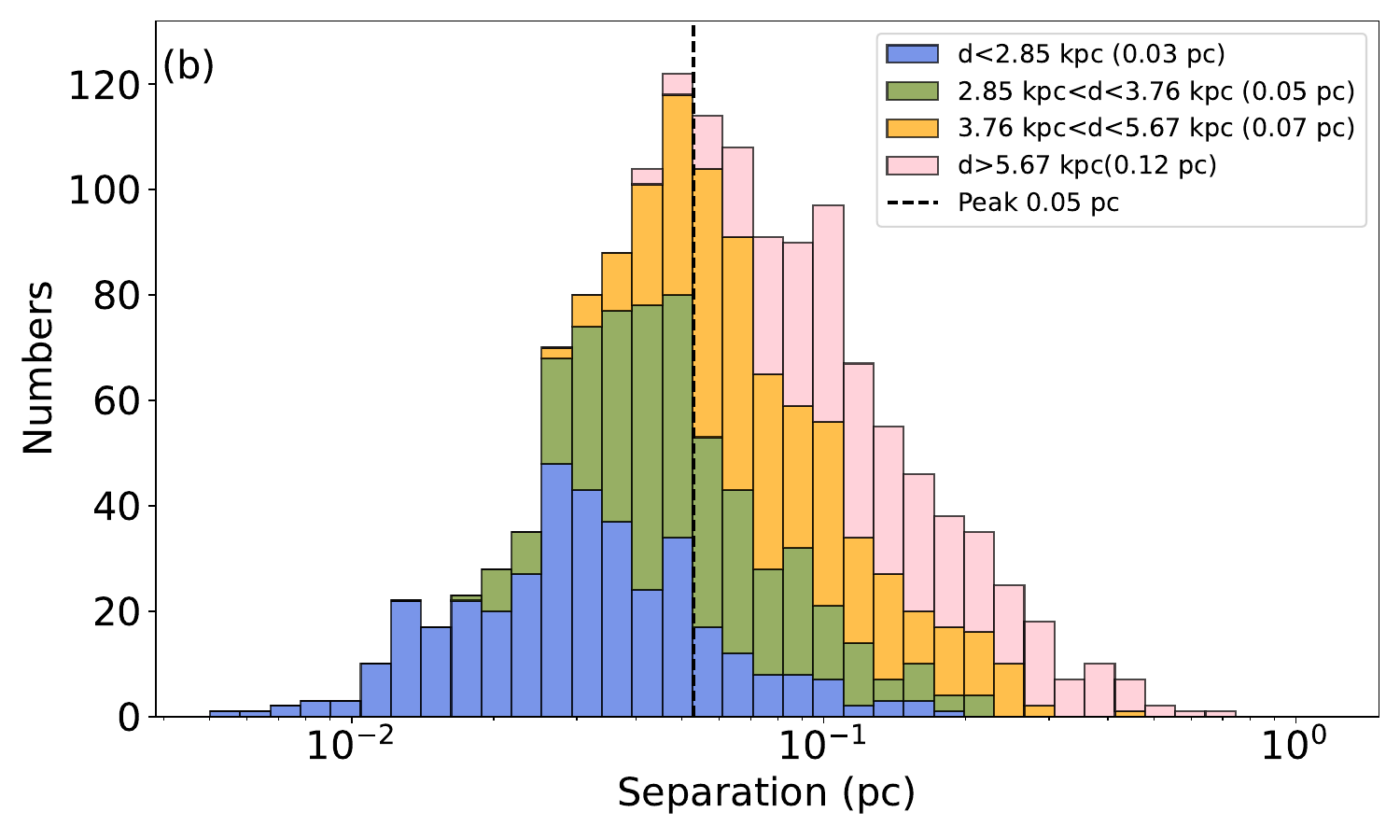}
    \caption{Distributions of core separation within all QUARKS protocluster clumps. To visualize the distance effect on the distribution,
    the clumps were devided into four quartile groups based on their measured distances: $\leq$2.85\,kpc, (2.85–3.76)\,kpc, (3.76–5.67)\,kpc, and $\geq$5.67\,kpc, with each quartile coded by a distinct color.
    The count of the corresponding groups is 36, 35, 35, and 35 protocluster clumps, respectively.
    Panel\,(a): distribution of the angular separation.  The number of core separation measurements for each group of protocluster clumps is 375, 363, 366, and 317, respectively, as shown in upper right legend.
    The black and red dashed lines indicate the peak of the distribution  ($\sim \rm 2.5$\arcsec), and the typical beam size ($\sim$1\,\arcsec) of our observations, respectively.
    Panel\,(b): distribution of the linear core separation. 
    The median value of the distribution for each group of protocluster clumps is 0.03\,pc, 0.05\,pc, 0.07\,pc and 0.12\,pc, respectively, as shown in upper right legend.
    The black dashed line indicates the peak of this distribution.} 
    \label{fig:sep}    
\end{figure}

\subsection{Core Separation}\label{sub:sec:res:separation}
Dense cores form through the fragmentation of their natal molecular clump. The Jeans length and mass are commonly employed to characterize this fragmentation process. However, as noted in \citet{2018A&A...617A.100B}, mass estimates are subject to significant uncertainties—not only from missing flux but also from assumptions about dust properties and temperatures (introducing uncertainties with factors of about 2, see Sect.\,\ref{sub:sec:res:physicalpara}). Consequently, core separation could serve as a more suitable proxy for analyzing fragmentation properties in clumps.

The minimum-spanning tree (MST; \citealt{1985MNRAS.216...17B,2019A&A...629A.135D}) method was used to determine the shortest separation between the cores within each clump (see Figure\,\ref{fig:MST_example} for two illustrative examples). This approach has been widely applied in studies of star-forming filaments, fibers, and clumps for core separation analysis \citep{2016ApJS..226...14W,2019ApJ...886..102S,2021A&A...646A..25Z,2022MNRAS.510.5009L,2023ApJ...950..148M,2024ApJS..270....9X,2024ApJ...976..241Y,2024ApJ...974...95I,2024ApJ...966..171M}.

Figure\,\ref{fig:sep}a presents the angular separation distribution for all cores in this study, spanning [0.9, 17.9]\,\arcsec\ (corresponding to $\rm [0.005, 0.72]\,pc$). The distribution peaks at 2.5\,\arcsec (black dashed line in Figure\,\ref{fig:sep}a), about twice the QUARKS TM2+ACA resolution ($\sim$1\,\arcsec), regardless of the source distance. This result suggests that all individual cores within each clump can be resolved by current angular resolution well.

In Figure\,\ref{fig:sep}b, the linear separation distribution exhibits a strong dependence on the source distance, with the median separation shifting toward higher values for farther distances (see the specific values in the legend of the figure, and also Figure\,\ref{fig:msep.dis}). This reflects an observational effect rather than a physical phenomenon, arising from the QUARKS survey's varying absolute mass sensitivity and linear resolution. For instance, the minimum detectable core mass increases with distance (Figure\,\ref{fig:m.dis}\,a), while the number of core detections (surface density of the core count) decreases systematically with distance (Figure\,\ref{fig:m.dis}\,b). These effects naturally lead to greater linear separations for more distant sources.

It should be noted that the measured core separations could represent lower limits due to projection effects, with actual values being potentially 1–2 times greater.

\subsection{Analysis of thermal Jeans length}\label{sub:sec:res:fragmentation}

\begin{figure}[ht!]
    \centering
    \includegraphics[angle=0, width=0.48\textwidth]{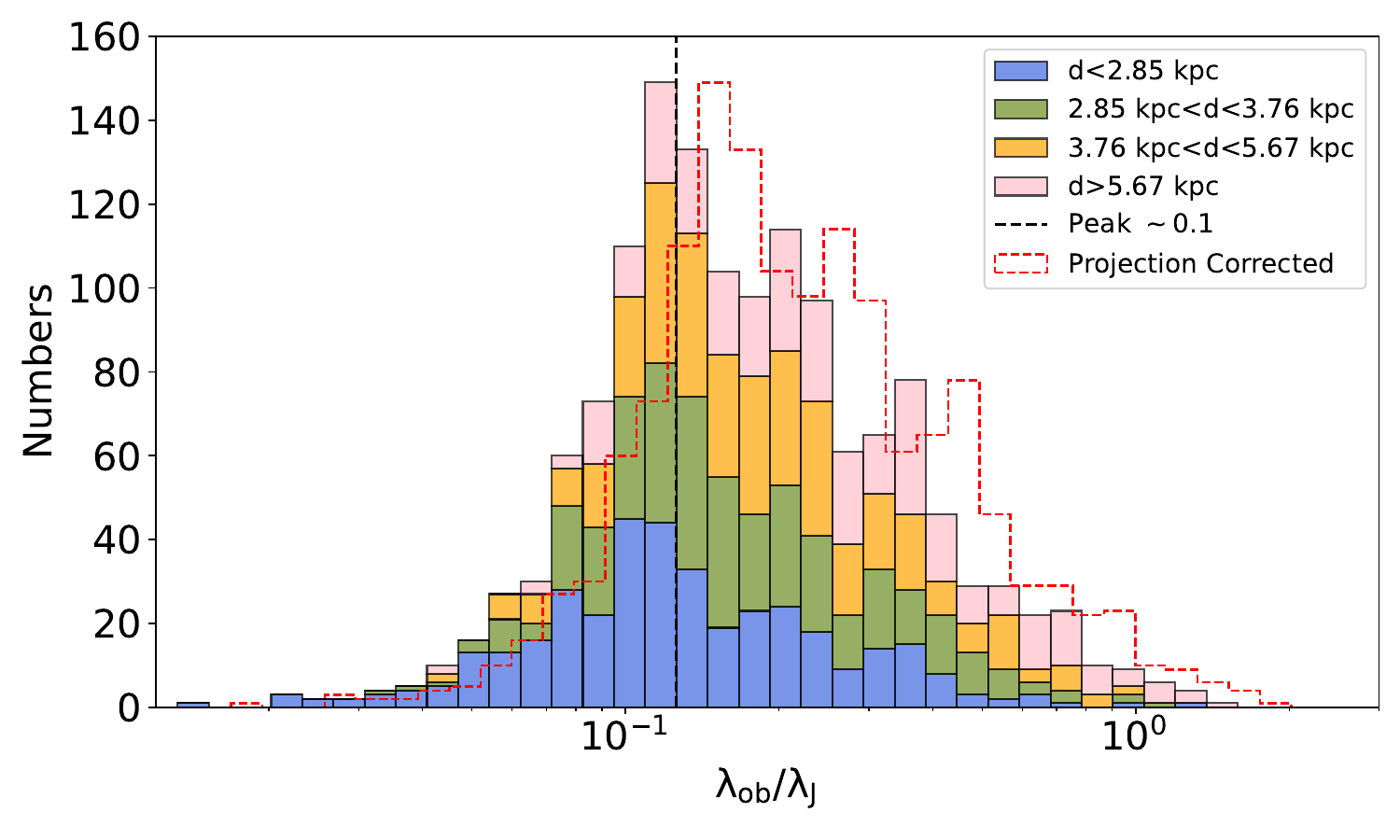} 
    \caption{Distribution of the ratio between the observed core separation to the predicted thermal Jeans length. The black dashed line shows the peak of the distribution. The red contour represents the distribution with the spatial projection correction applied (see text). 
    } 
    \label{fig:Jeans_sep}    
\end{figure}

To investigate the clump-to-core fragmentation using the QUARKS TM2+ACA continuum data, we conducted the comparison between the observed linear separation of the cores and the predicted thermal Jeans length within their natal clumps.
Using physical parameters from Table A1 in \cite{2020MNRAS.496.2790L}, we find that the clump number densities range from $\sim\rm 10^3-10^5\,cm^{-3}$ and the dust temperatures vary between 19.5 and 45.5 K. These conditions yield Jeans lengths ($\lambda_{\rm J}=c_s(\pi/G\rho)^{1/2}$) of $\rm\sim [0.07, 1.1]\,pc$ for all clumps of the QUARKS survey.

The ratio of the observed core separation to the thermal Jeans length ($\lambda_{\rm obs}/\lambda_{\rm J}$) is shown in Figure\,\ref{fig:Jeans_sep}, revealing a distribution apparently independent of the source distance. Both $\lambda_{\rm obs}$ and $\lambda_{\rm J}$ exhibit positive correlations with source distance, as demonstrated by Spearman's rank correlation tests\,\footnote{The Spearman's rank correlation test is suitable for small sample sizes and does not require the variables to follow a Gaussian (normal) distribution. A $\rho_s$ closer to +1 or -1 indicates a stronger positive or negative monotonic relationship, respectively, and the value of 0 suggests no correlation. If p-value<0.01, the result is considered statistically significant.}: $\rho_s \sim 0.9$ ($p\ll0.01$) for the $\lambda_{\rm obs}$ in Figure\,\ref{fig:msep.dis}, and $\rho_s \sim 0.6$ ($p\ll0.01$) for the $\lambda_{\rm J}$ in Figure\,\ref{fig:s_j.dis}a.  
However, the $\lambda_{\rm obs}/\lambda_{\rm J}$ ratio can effectively mitigate (or reduce) the effects related to the source distance, as illustrated in Figure\,\ref{fig:s_j.dis}\,b ($\rho_s\sim0.4$ and $p\ll 0.01$).
Almost all ratios ($\sim99\%$, 1407/1421) are below 1, peaking at $\sim\rm 0.1$ (black dashed line in Figure\,\ref{fig:Jeans_sep}). After applying the correction for projection effects (consider the factor of $4/\pi$, e.g., \citealt{2024ApJ...974...95I}), the red histogram in Figure\,\ref{fig:Jeans_sep} shows that $\sim98\%$ (1391/1421) of the ratios remain below unity, with a peak at $\sim\rm 0.2$. This result indicates that the massive protocluster clumps studied here fragment with core separations significantly smaller than the predicted thermal Jeans lengths. This agrees with recent studies showing observed core separations comparable to or less than the thermal Jeans length both in IR-dark clouds \citep{2019ApJ...886..102S,2024ApJ...966..171M}, and in more evolved IR-bright star-forming clumps \citep{2015MNRAS.453.3785P,2018ApJ...855...24P,2018A&A...617A.100B,2021A&A...646A..25Z,2024ApJ...974...95I}.

Interestingly, a secondary peak in Figure\,\ref{fig:Jeans_sep} appears at $\lambda_{\rm obs}/\lambda_{\rm J}\sim 0.3$.
A similar bimodal distribution of core separations has also been observed \citep{2016A&A...587A..47T,2018ApJ...855...24P,2019ApJ...886...36S,2021A&A...646A..25Z}, which could be attributed to hierarchical fragmentation occurring at two distinct scales within a clump. Note that the secondary peak is much weaker than the primary one, and thus should be treated with caution.

\section{Discussion}\label{sec:dsicussion}

\subsection{Fragmentation of Protocluster clumps}\label{sec:frg}
Recent high-angular-resolution ALMA observations suggest that fragmentation of both IR-dark and IR-bright clumps could follow the thermal Jeans fragmentation mechanism. For instance, in the ALMA-ASHES survey targeting tens of IRDC clumps, core separations align with the predicted thermal Jeans length, while core masses fall below the thermal Jeans mass \citep{2019ApJ...886..102S,2024ApJ...966..171M}. Similar results were observed in evolved protocluster clumps \citep{2015MNRAS.453.3785P,2018A&A...617A.100B,2024MNRAS.534.3832D} and those associated H{\sc ii} region environments \citep{2021A&A...646A..25Z}. Our QUARKS sample, which includes IR-bright massive protocluster clumps more evolved than the ASHES sample \citep{2024RAA....24f5011X}, with most hosting embedded UCH{\sc ii} sources \citep{2021MNRAS.505.2801L,2023MNRAS.520..322Z,2024MNRAS.533.4234Z}, shows core separations significantly smaller than the corresponding thermal Jeans length ($\lambda_{\rm obs}/\lambda_J \sim 0.2$, accounting for projection effects). This suggests that the IR-bright protocluster clumps in our study fragment primarily through thermal Jeans fragmentation.

Moreover, the peak of the core separation distribution at $\sim$0.05\,pc (Figure\,\ref{fig:sep}\,b) agrees with that observed in other IR-bright massive clumps ($\sim$0.04\,pc; see Figure\,6 of \citealt{2024ApJS..270....9X}) but is significantly less than those measured in IR-dark massive clumps ($\sim$0.1\,pc; \citealt{2019ApJ...886..102S,2023ApJ...950..148M}). This difference in typical core separations between IR-dark and IR-bright clumps could naturally lead to the observed low ratio of $\lambda_{\rm obs}/\lambda_J \sim 0.2$. This result could be linked to the evolutionary stage of the natal clumps, as suggested by recent studies \citep{2018A&A...617A.100B,2023MNRAS.520.2306T}. Specifically, dense cores could initially form via thermal Jeans fragmentation during the IR-dark stage with slightly larger separations, and then evolve toward less separations due to the persistent global gravitational collapse and contraction over time \citep{2016A&A...587A..47T,2024ApJS..270....9X,2024MNRAS.534.3832D}. However, when using the luminosity-to-mass ratio ($L/M$) as an evolutionary indicator \egcite{2021MNRAS.505.2801L} for the QUARKS protocluster clumps, we do not find a clear correlation between the $L/M$ and core separation, which is not shown here. This lack of correlation could stem from a limited dynamical range of evolutionary phases in the QUARKS sample, which includes IR-bright stages only. Additionally, as protocluster clumps evolve, they could generate new fragments, leading to a reduction in core separations over time, as reported in recent studies using ALMAGAL survey data at $\sim$1000\,AU resolution  \citep{2025arXiv250305663C}.  

An alternative explanation for the observed $\lambda_{\rm obs}/\lambda_{\rm J} \sim 0.2$ significantly below unity could be due to the existence of a hierarchical fragmentation mode within the clumps. Through this mode, massive protocluster clumps initially fragment into subclumps, which subsequently break down hierarchically into smaller-scale dense cores \citep{2024ApJ...966..171M}. As a first approximation, considering that the number densities of subclumps in the QUARKS survey to be those derived from the ACA compact sources (i.e., $\sim 10^5-10^6\,\mathrm{cm}^{-3}$, \citealt{2024RAA....24f5011X}), the Jeans lengths within subclumps would decrease by a factor of $\sim 3-10$ with respect to those within their larger-scale natal clumps as analyzed in Sect.\,\ref{sub:sec:res:fragmentation}. This would shift the peak of the $\lambda_{\rm obs}/\lambda_{\rm J}$ distribution in Figure\,\ref{fig:Jeans_sep} to $\sim 0.3-1$, which still agrees with the thermal Jeans fragmentation mechanism well. Future detailed investigations into hierarchical fragmentation involving (sub)clump scales based on the QUARKS survey are required (W.Y. Jiao et al., in preparation). Overall, regardless of the potential evolution of the core separation with the clump evolutionary stage and likely existence of a hierarchical fragmentation mode, the observed low ratio of $\lambda_{\rm obs}/\lambda_J \ll 1$  indicates that the massive protocluster clumps investigated here are undergoing thermal Jeans fragmentation.

Unlike our findings, some studies found that in IR-dark clumps there are dense core masses exceeding the thermal Jeans mass by at least an order of magnitude, consistent with expectations for turbulent Jeans fragmentation of their natal clumps \citep{2009ApJ...696..268Z,2011ApJ...733...26Z,2013ApJ...773..123S,2014MNRAS.439.3275W}. As noted in other studies, this discrepancy may arise either from observational limitations—such as insufficient mass sensitivity ($>2M_\odot$) and spatial resolution ($>5000$\,au) \citep{2015MNRAS.453.3785P}—or from the fact that low-mass cores (or fragments) not forming yet at such early evolutionary stages \citep{2014MNRAS.439.3275W,2015ApJ...804..141Z}. Alternative explanations for this discrepancy include steep initial density profiles (e.g., $\rho \propto r^{-1.5}$ or $\rho \propto r^{-2}$), which favor the formation of central massive cores \citep{2011MNRAS.413.2741G,2014ApJ...785...42P}. Additionally, stronger magnetic fields have been proposed to suppress fragmentation and promote massive core formation \citep{2007ApJ...661..262D,2011ApJ...742L...9C,2011A&A...530A.118P,2016A&A...593L..14F,2022A&A...668A.147H}. 
However, recent studies have argued that strong magnetic fields do not reverse the thermal Jeans fragmentation process but instead act as a modulating factor \citep[e.g.,][]{2021ApJ...912..159P,Klos2025MNRAS.539.2307K} that facilitates the formation of a few high-mass stars \citep[e.g.,][]{2024ApJ...972L...6S,2025ApJ...980...87S}. 
In the future, systematic investigations using the QUARKS sample will be essential to clarify the influence of these factors on clump fragmentation.

\begin{figure}[ht!]
    \centering
    \includegraphics[angle=0, width=0.4\textwidth]{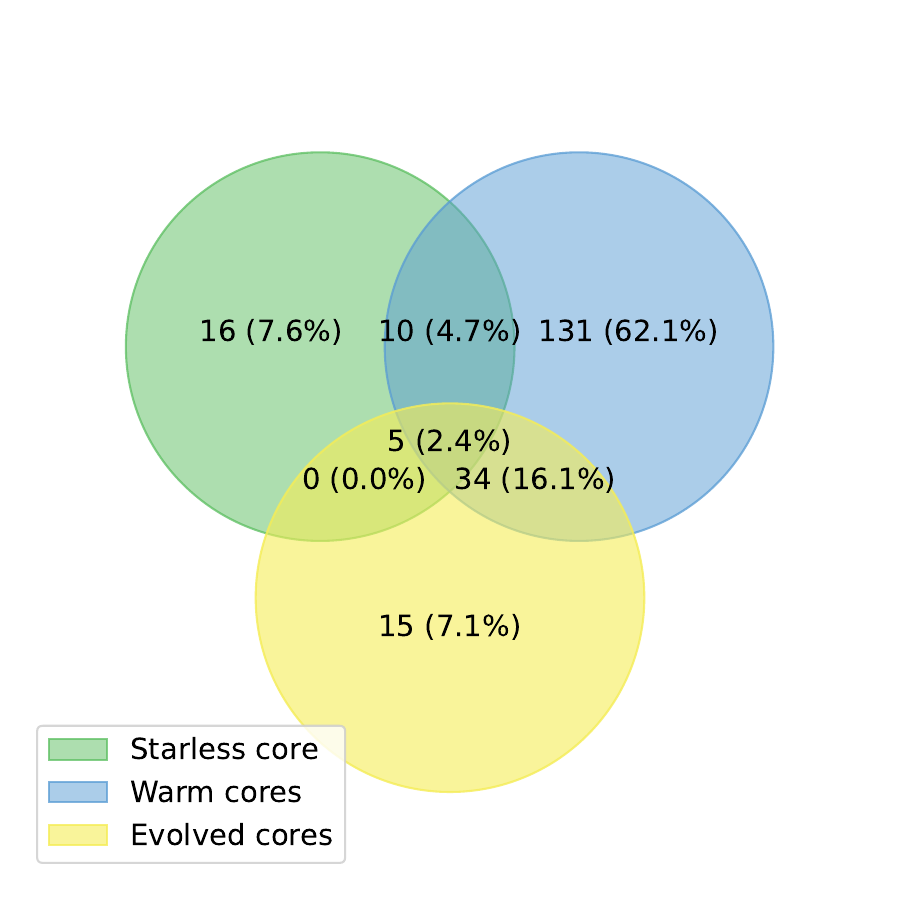} 
    \caption{Venn diagram of cross-matched ATOMS ``$unknown$" sources with QUARKS TM2+ACA cores at different evolutionary stages. The number in each circle represents the cross-matched result.}
    \label{fig:venn}    
\end{figure}

\subsection{Comparison with ATOMS ``$unknown$" source catalog}
In the ATOMS survey, \cite{2021MNRAS.505.2801L} identified 280 compact fragments lacking signatures of HMCs or UCH{\sc ii} regions, designating them as ATOMS ``$unknown$" sources. 
These sources exhibit median properties of $2.4\times10^{-2}$\,pc radius in an interquartile range (IQR) of [1.4, 4.6]$\times10^{-2}$\,pc, 63\,M$_{\odot}$ mass in an IQR of [20, 320]\,M$_{\odot}$, $9\times10^{6}$\,\pcmcu\ number density in an IQR of [4, 17]$\times10^{6}$\,\pcmcu\, and 7.2\,g~\cmcm\ mass surface density in an IQR of [4.3, 14.2]\,g~\cmcm\,.
Due to these characteristics of high mass and density, \cite{2021MNRAS.505.2801L} suggested that the ``$unknown$" sources might represent precursors to high-mass star formation. We cross-matched the ATOMS ``$unknown$" sources sample with the 1562 QUARKS TM2+ACA cores from this study (Figure\,\ref{fig:venn}). Of the 280 "$unknown$" sources, 69 lie outside the QUARKS survey field, while the remaining 211 correspond to 464 QUARKS cores. This suggests that each ``$unknown$" source could fragment into at least two QUARKS cores at $\sim$1\,\arcsec\ resolution.
The majority of the ``$unknown$" sources sample (75$\%$, 157/211) harbor candidate starless and/or warm cores, while 25$\%$ (54/211) contain evolved cores. This indicates that the catalogue of ATOMS ``$unknown$" sources is a suitable sample to investigate initial conditions of high-mass star formation. Particularly, 16 of these ATOMS ``$unknown$" sources are uniquely linked to candidate starless cores within the QUARKS TM2+ACA catalogue (see Figure\,\ref{fig:venn}), suggesting their potential as prime candidates in the initial phases of high-mass star formation core evolution and thus highlighting the strong need for detailed follow-up studies.

\subsection{ Search for candidate high-mass starless cores and its implication for high-mass star formation}\label{sec:hpc}

Competitive accretion-type models predict that a massive clump fragments into numerous low-mass cores that subsequently promote high-mass star formation through competitive gas accretion \citep{2001MNRAS.323..785B} within the clump and even beyond \citep{2019MNRAS.490.3061V,2020ApJ...900...82P}. In contrast, the existence of individual high-mass starless cores is crucial for validating turbulent core accretion-type models \citep{2003ApJ...585..850M}. To date, around a handful of candidate high-mass starless cores have been identified only \citep{2014MNRAS.439.3275W,2017ApJ...849...25L,2023A&A...675A..53B,2023A&A...674A..75N,2024ApJ...961L..35M,2025arXiv250209426V}.

In this study, we classified 127 candidate starless cores. Assuming a star formation efficiency of 50\% as adopted in \citet{2018A&A...618L...5N} and \cite{2025arXiv250209426V}, only two candidates (with $M_{\rm core} \geq 16\,M_{\odot}$ at $T_{\rm d} = 20\,\rm K$) could potentially form high-mass stars independently. 
We highlight these two cores in boldface in Table\,\ref{table:phypre}, which have radii of 5000–7000\,au and masses of 17–21\,M$_{\odot}$.
The virial parameter ($\alpha_{\rm vir}=5\sigma_{\rm eff}^2R_{c}/\rm G{\it M_{\rm c}}$, \citealt{1992ApJ...395..140B}) is widely used to evaluate the dynamical state of a core (e.g., \citealt{2021MNRAS.505.2801L,2021MNRAS.503.4601B,2024ApJS..270....9X,2025arXiv250209426V}). 
Considering typical total velocity dispersions of high-mass prestellar cores in the literature ($\sim\rm0.4-0.8\,km\,s^{-1}$, e.g., \citealt{2023A&A...675A..53B,2024ApJ...961L..35M}), we find $\alpha_{\rm vir} < 2$ for the two candidates studied here. This indicates that they are gravitationally bound and could be therefore  candidate prestellar cores, precursors of high-mass star formation.
Line spectra of the two candidate cores and detection maps of CO outflows within their natal clumps are shown in Figures\,\ref{fig:pscline} and \ref{fig:cooutflow}, respectively.

It is worth noting that the number of high-mass starless candidates identified here likely represents a lower limit.
First, if a lower dust temperature was considered (10–15\,K, e.g., \citealt{2018ARA&A..56...41M,2023A&A...675A..53B,2023ApJ...950..148M,2024ApJ...961L..35M}),
this would shift the masses of six cores (i.e., $ 8-16\,M_{\odot}$ at 20\,K) to values greater than 16\,\msun,  potentially adding additional high-mass starless candidates. Second, QUARKS' high-angular-resolution observations at 1.3\,mm  primarily trace dense kernels of cores, likely filtering out significantly their envelope mass. In contrast, the ATOMS survey observations at 3\,mm at a lower-angular-resolution retain relatively more extended envelope emission \citep{2024ApJ...961L..35M}. 
As illustrated in Figure\,\ref{fig:atomsmass:quarksmass}, the ATOMS ``unknown" sources exhibit significantly larger median size and mass than the QUARKS candidate starless cores studied here, with the difference on the median size and mass by factors of $\sim$3 and $\sim$50, respectively.
Note that only 26 ATOMS ``unknown" sources that are cross-matched with the QUARKS candidate starless cores (see Figure\,\ref{fig:venn}) are depicted in Figure\,\ref{fig:atomsmass:quarksmass}.
Future studies combining QUARKS and ATOMS data are therefore needed to better evaluate the mass threshold for potential high-mass starless candidates. Furthermore,  there could be potential high-mass starless cores in our warm core catalog, arising from those $\sim 5\%$ ($\sim$50/971) cores with $ M_{\rm core}\ge8\,M_{\odot}$ at $T\rm_d=30\,K$ (see Appendix\,\ref{Appa}) that are not associated with any star-forming signatures (e.g., outflows).

Regardless of the precise candidate count, we find that 90\% of candidate starless cores ($M_{\rm core} \leq 8\,M_{\odot}$) lack sufficient mass for monolithic high-mass star formation. However, surrounding filaments and streamers may channel gas mass onto these cores \citep{2022MNRAS.516.1983S,2022MNRAS.514.6038Z,2023ApJ...953...40Y,2023MNRAS.520.3259X}, enabling future high-mass star formation \citep{2018ARA&A..56...41M} as depicted in the competitive accretion model. Our results therefore require competitive accretion-like models where clumps first fragment into low-mass cores and high-mass stars primarily grow through multi-scale dynamical mass accretion from beyond their immediate gas reservoir, with rare cases forming via monolithic collapse with associated mass accretion onto protostars from the envelope of their natal massive starless cores.

\section{Summary}\label{sec:summary}
The QUARKS survey observed 139 IR-bright massive star-forming protocluster clumps across 156 pointings using ALMA band\,6 ($\sim\rm1.3\,mm$). These observations employed three configurations of the ALMA 12-m array: C-5 (TM1, $\sim0.3$\,\arcsec), C-2 (TM2, $\sim1$\,\arcsec), and ACA 7-m array ($\sim5$\,\arcsec). This study details the imaging procedures to generate the continuum and line datasets from the combined TM2+ACA observations. Employing self-calibration in both continuum and line imaging, the final reduced TM2+ACA combined data achieved an average noise level of $\sim\rm 0.6\,mJy\,beam^{-1}$ with a synthesized beam of $\sim1.3$\,\arcsec$\times1.1$\,\arcsec\ for continuum, and  an average noise level of $\sim \rm 10\,mJy\,beam^{-1}$ per channel at a velocity resolution of $\sim\rm 1.3\,km\,s^{-1}$ and a synthesized beam of $\sim 1.4$\,\arcsec$\times1.1$\,\arcsec\ for spectral lines.

Furthermore, 1562 compact cores were extracted from the TM2+ACA combined continuum data using the {\it getsf} algorithm. Using these cores, we conducted a systematic investigation into clump-to-core fragmentation  and a preliminary search for candidate high-mass starless cores. The major findings are summarized below.

Based on associated star-forming signatures, namely the richness of molecular transitions ($N_{\rm line}$), outflows, and/or ionized gas tracers, we categorized the identified cores into three groups: 127 candidate starless cores, 971 candidate warm cores (WCs), and 464 candidate evolved cores. The evolved candidates consist of three subcategories: 314 candidate hot molecular cores (HMCs), 63 candidate UCH{\sc ii}-{\it h} cores (UCH{\sc ii}-{\it h}) with high $N_{\rm line}$, and 87 candidate UCH{\sc ii}-{\it l} cores (UCH{\sc ii}-{\it l}) with low $N_{\rm line}$. The median size, mass, and surface density were estimated as follows: 2900\,au, 1.6\,\msun, and $\rm 0.6\,g\,cm^{-2}$ for candidate starless cores; 2600\,au, 2.0\,\msun, and $\rm 0.7\,g\,cm^{-2}$ for candidate WCs; 2700\,au, 3.7\,\msun, and 1.4\,g~cm$^{-2}$ for evolved cores.

Employing the MST method, we quantified the core separations within each protocluster clump. The angular separations of all cores range from 0.9\,\arcsec to 17.9\,\arcsec, peaking at 2.5\,\arcsec, approximately twice the TM2+ACA resolution ($\sim$1\,\arcsec), indicating that the studied cores are spatially resolved well. In contrast, the distribution of the linear core separation, spanning [0.005, 0.72]\,pc and peaking at $\sim 0.05$\,pc, shows a strong positive correlation with the source distance, with the core separation increasing towards a far distance.

Moreover, the observed linear core separations ($\lambda_{\rm obs}$) within each clump are much less than the related theoretical thermal Jeans length ($\lambda_{\rm J}$), yielding a ratio of $\lambda_{\rm obs}/\lambda_{\rm J}\sim0.2$ when accounting for projection effects. This finding indicates that the massive protocluster clumps in this study are undergoing thermal Jeans fragmentation. We further suggest that the observed low ratio of $\lambda_{\rm obs}/\lambda_{\rm J}\ll 1$ could stem from either the potential evolution of core separation with the clump's evolutionary stage or the likely presence of hierarchical fragmentation within clumps.

The IR-bright environment of massive protocluster clumps, characterized by typically higher gas temperatures compared to IR-dark regions, may foster the formation of high-mass starless cores. Within 127 candidate starless cores revealed in QUARKS TM2+ACA continuum emission, we identified two cores exceeding 16\,\msun\ as high-mass candidates, possessing radii of 3000–8000\,au and masses of 17–21\,M$_{\odot}$. The number of such candidates identified here represents a lower limit due to the inherent mass estimate uncertainty of the cores, which warrants further investigation for starless cores by integrating the ATOMS and QUARKS datasets. Irrespective of the exact candidate count, we find that 90\% of candidate starless cores ($M_{\rm core} \leq 8\,M_{\odot}$) do not possess sufficient mass for monolithic high-mass star formation. The scarcity of such high-mass starless candidates therefore suggests that competitive accretion-like models are more applicable than turbulent core accretion-like models in high-mass star formation within the IR-bright protocluster clumps studied here.

\section*{Acknowledgements}
This work has been supported by the National Key R\&D Program of China (No.\,2022YFA1603101), the National Natural Science Foundation of China (NSFC) through grants No.12073061, No.12122307, and No.12033005.
H.-L. Liu is supported by Yunnan Fundamental Research Project (grant No.\,202301AT070118, 202401AS070121), and by Xingdian Talent Support Plan--Youth Project. 
D.-T. Yang is supported by the Scientific Research Fund Project of Yunnan Education Department (Project ID:\,2025Y0106, KC-24248416).
Tie Liu acknowledges the supports by the PIFI program of Chinese Academy of Sciences through grant No. 2025PG0009, and the Tianchi Talent Program of Xinjiang Uygur Autonomous Region. 
SZ gratefully acknowledges support by the CAS-ANID project CAS220003.
X. Liu has also been supported by CPSF No.\,2022M723278.
P.S. was partially supported by a Grant-in-Aid for Scientific Research (KAKENHI Number JP22H01271 and JP23H01221) of the Japan Society for the Promotion of Science (JSPS).
AS gratefully acknowledges support by the Fondecyt Regular (project code 1220610), and ANID BASAL project FB210003. 
GG gratefully acknowledges support by the ANID BASAL project FB210003.
K.T. is supported by JSPS KAKENHI grant Nos.\,21H01142, 24K17096, and 24H00252.
L.B. gratefully acknowledges support by the ANID BASAL project FB210003.
A.P. acknowledges financial support from the UNAM-PAPIIT IG100223 grant, the Sistema Nacional de Investigadores of SECIHTI, and from the SECIHTI project number 86372 of the `Ciencia de Frontera 2019’ program, entitled `Citlalc\'oatl: A multiscale study at the new frontier of the formation and early evolution of stars and planetary systems’, M\'exico.
T. Z acknowledges the National Natural Science Foundation of China (Grant No. 12373026), the Leading Innovation and Entrepreneurship Team of Zhejiang Province of China (Grant No. 2023R01008), and the China Postdoctoral Science Foundation (Grant No. 2023TQ0330).
AH thanks the support by the S. N. Bose National Centre for Basic Sciences under the Department of Science and Technology, Govt. of India and the CSIR-HRDG, Govt. of India for the funding of the fellowship.
This work was performed in part at the Jet Propulsion Laboratory, California Institute of Technology, under contract with the National Aeronautics and Space Administration.
This paper makes use of the following ALMA data: ADS/JAO.ALMA\#2021.1.00095.S. 

\clearpage
\appendix
\section{Physical parameters of the QUARKS TM2+ACA dense cores}\label{Appa}
\begin{figure}[ht!]
    \centering
    \includegraphics[angle=0, width=0.45\textwidth]{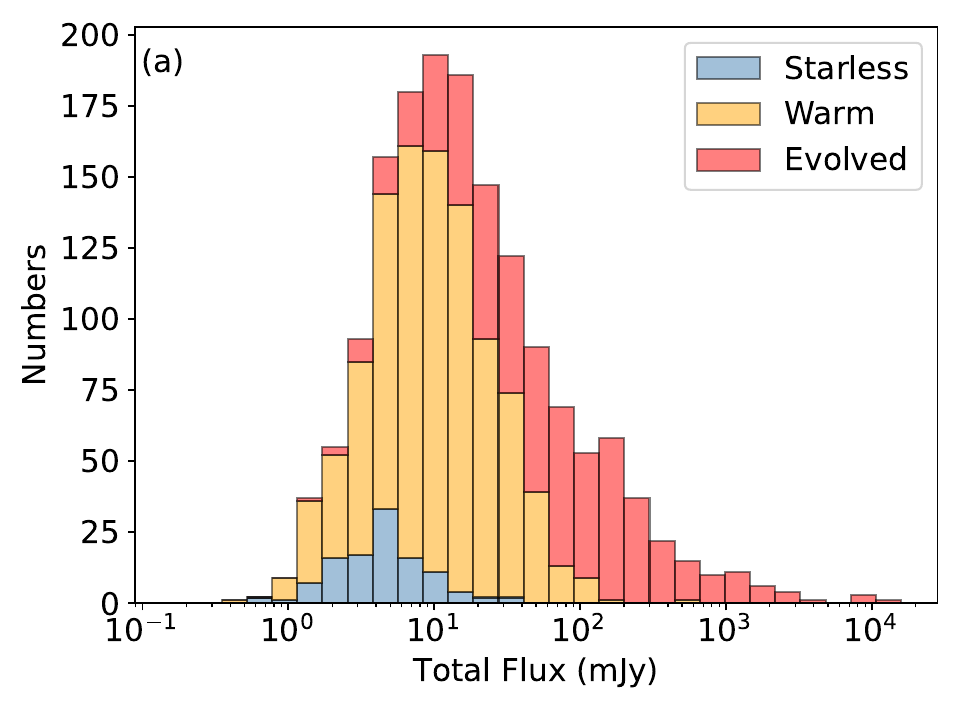} 
    \includegraphics[angle=0, width=0.45\textwidth]{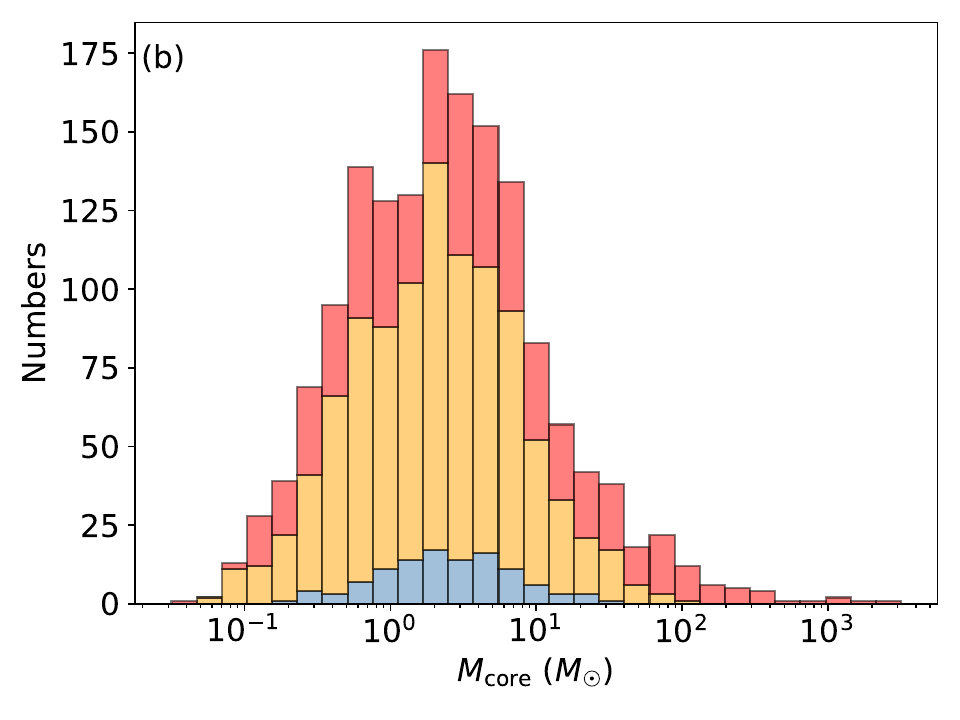} 
    \includegraphics[angle=0, width=0.45\textwidth]{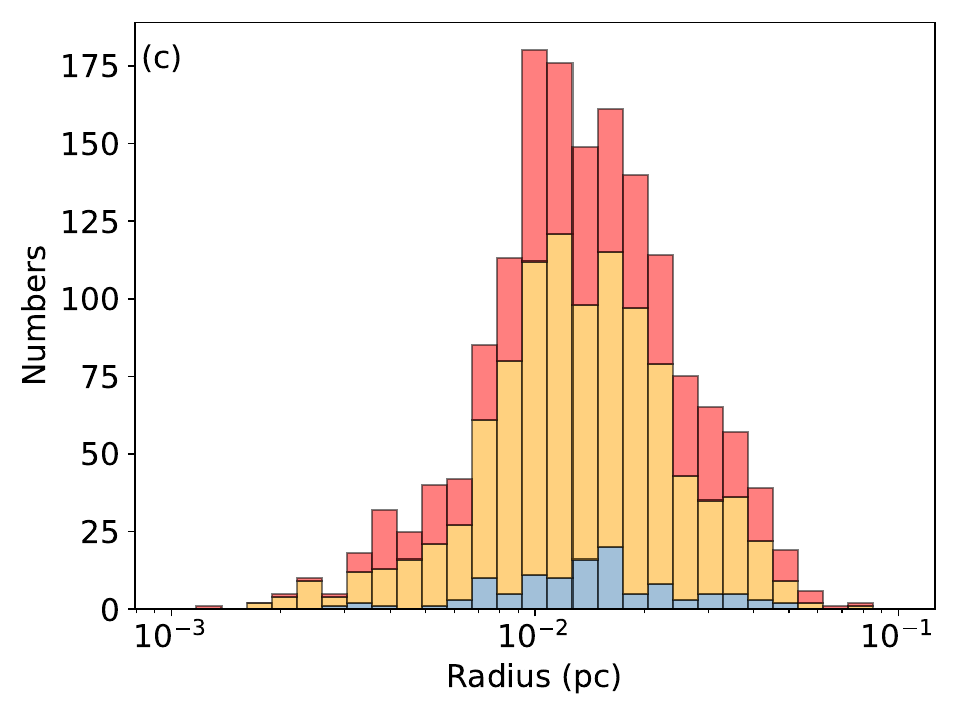} 
    \includegraphics[angle=0, width=0.45\textwidth]{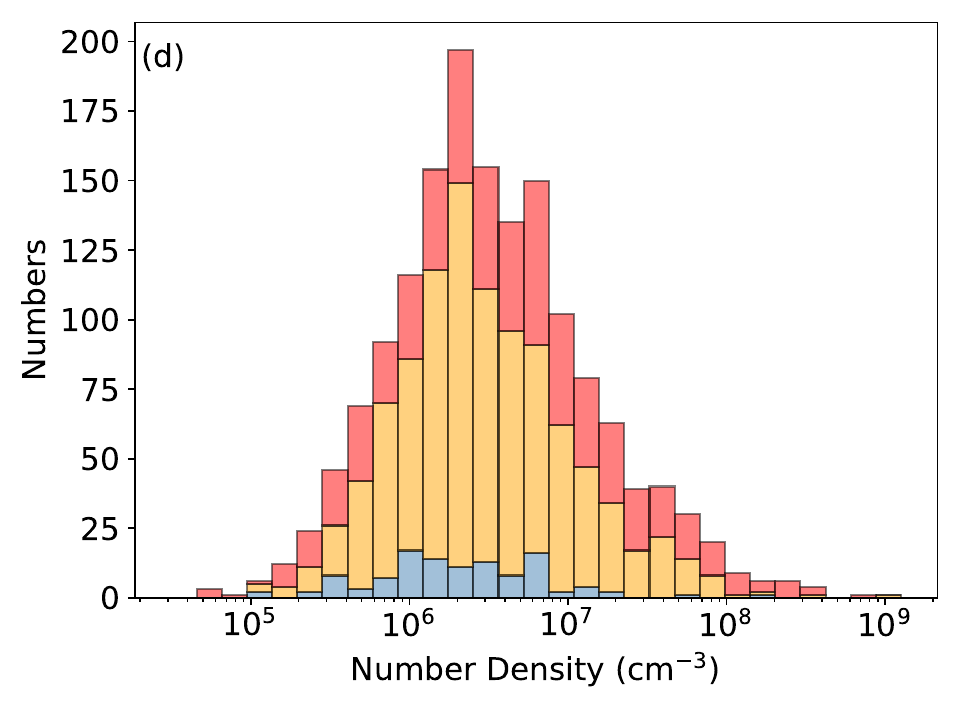} 
    \includegraphics[angle=0, width=0.45\textwidth]{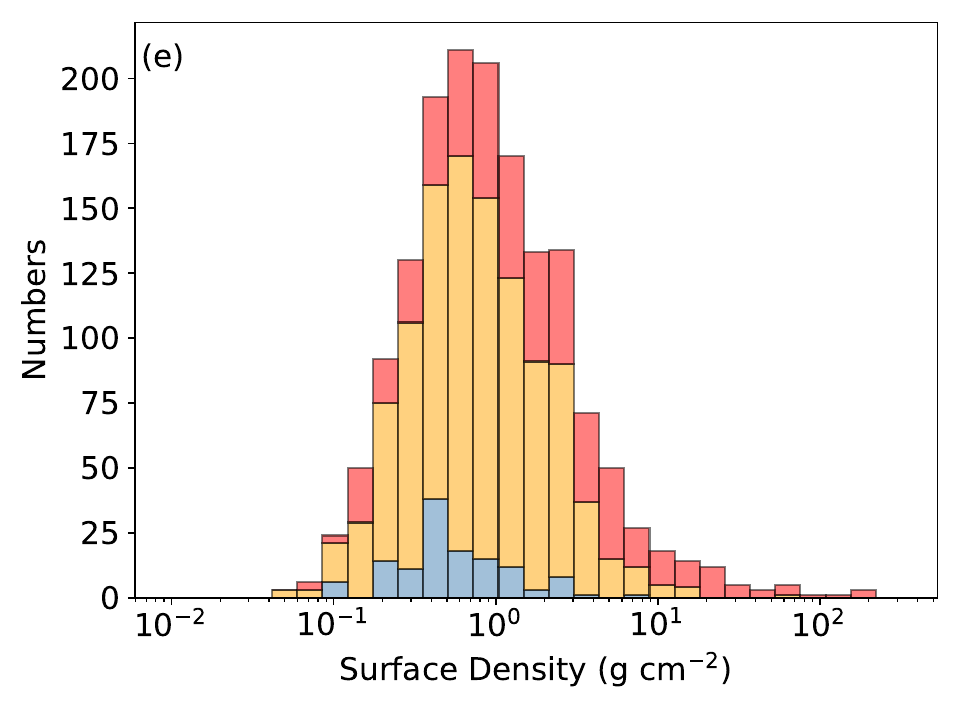} 
    \caption{Distribution of physical parameters of the QUARKS TM2+ACA dense cores against  evolutionary stage. Panels\,a-e present the parameter of total flux, mass, radius, number density, and surface density, respectively.} 
    \label{fig:corephysical}    
\end{figure}

Derived parameters for candidate starless cores are listed in Table\,\ref{table:phypre}, and those for candidate warm and evolved cores are presented in Table\,\ref{table:phywarm}. Figure\,\ref{fig:corephysical} presents the distribution of physical parameters for dense cores across different evolutionary stages. While the total flux of cores generally increases with evolutionary stage, mass, radius, and number density do not show clear trends. Note that 27 cores exceeding 100\,\msun\ (Fig\,\ref{fig:corephysical}b) are evolved cores (11 in HMCs and 15 in UCH{\sc ii} cores), which require careful consideration due to their association with ionized gas emission. From the cross-match between these cores and MeerKAT 1.3\,GHz radio continuum arising from free-free emission, we find that except for the UCH{\sc ii} cores, almost all HMCs in question lie in extended ionized gas emission. The 1.3\,mm flux of these 27 cores could be significantly affected by free-free emission, potentially leading to mass overestimation.

\startlongtable
\begin{deluxetable*}{ccccccccccc}\label{table:phypre}
\tabletypesize{\footnotesize}
% \tablewidth{0pt} 
\tablecaption{Derived parameters for candidate starless cores.} 
% \tablecomments{}
% \centering
\tablehead{
\colhead{Source} & \colhead{Core} & \colhead{RA\,(IRCS)} & \colhead{Dec\,(IRCS)}& \colhead{$R$}& \colhead{$N_{\rm line}^{(\it {a})}$}& \colhead{$\rm Temp.^{(\it b)}$}& \colhead{$\rm Mass_{upper}^{(c)}$}& \colhead{$\rm Mass_{lower}^{(c)}$}& \colhead{$n_{\rm H_2(20\,K)}^{\it (d)}$}& \colhead{$\rm \Sigma_{(20\,K)}^{\it (d)}$}\\
&& \colhead{(h:m:s)}  & \colhead{(d:m:s)}& \colhead{(au)} & &(K)& \colhead{(\msun)}& \colhead{(\msun)} & $(\rm\times 10^6\,cm^{-3})$ & $(\rm g\,cm^{-2})$ \\ }
\colnumbers 
\startdata 
I08448-4343 & 9 & 8h46m31.99s & -43d54m40.07s & 535 & 5 & 25 & 0.3 & 0.23 & 59.3 & 3.0 \\
I08448-4343 & 20 & 8h46m33.89s & -43d54m27.39s & 744 & 2 & 25 & 0.21 & 0.16 & 15.4 & 1.1 \\
I09002-4732 & 8 & 9h01m54.91s & -47d44m01.87s & 628 & 6 & 39 & 1.21 & 0.54 & 147.9 & 8.7 \\
I09094-4803 & 6 & 9h11m08.68s & -48d16m04.62s & 8752 & 3 & 23 & 7.01 & 5.86 & 0.3 & 0.3 \\
I09094-4803 & 7 & 9h11m08.87s & -48d16m04.12s & 6003 & 3 & 23 & 2.9 & 2.43 & 0.4 & 0.2 \\
I09094-4803 & 9 & 9h11m08.61s & -48d15m51.44s & 4234 & 4 & 23 & 2.26 & 1.89 & 0.9 & 0.4 \\
I12572-6316$\_$2 & 3 & 13h00m28.74s & -63d32m29.79s & 7832 & 2 & 21 & 10.93 & 10.26 & 0.7 & 0.5 \\
I12572-6316$\_$2 & 4 & 13h00m28.11s & -63d32m38.47s & 10540 & 4 & 21 & 12.76 & 11.98 & 0.3 & 0.3 \\
I13079-6218 & 5 & 13h11m08.42s & -62d34m44.89s & 2070 & 3 & 28 & 4.71 & 3.09 & 16.1 & 3.1 \\
I13079-6218 & 6 & 13h11m09.29s & -62d34m41.35s & 2009 & 5 & 28 & 3.33 & 2.18 & 12.4 & 2.3 \\
I13111-6228 & 11 & 13h14m28.01s & -62d44m27.77s & 1570 & 4 & 30 & 0.49 & 0.3 & 3.8 & 0.6 \\
I13134-6242 & 3 & 13h16m42.27s & -62d58m26.40s & 1853 & 4 & 29 & 1.33 & 0.84 & 6.3 & 1.1 \\
I13140-6226 & 12 & 13h17m16.14s & -62d42m34.20s & 3419 & 5 & 22 & 1.57 & 1.39 & 1.2 & 0.4 \\
I13291-6229 & 10 & 13h32m30.46s & -62d45m07.80s & 3787 & 2 & 28 & 1.32 & 0.86 & 0.7 & 0.3 \\
I13291-6229 & 12 & 13h32m30.53s & -62d45m10.99s & 3624 & 2 & 28 & 0.71 & 0.47 & 0.5 & 0.2 \\
I13291-6229 & 13 & 13h32m35.50s & -62d45m33.33s & 2479 & 2 & 28 & 0.69 & 0.45 & 1.4 & 0.3 \\
I13295-6152 & 7 & 13h32m53.25s & -62d07m56.08s & 3310 & 2 & 19 & 1.24 &  1.16 & 1.0 & 0.3 \\
I13295-6152 & 10 & 13h32m53.39s & -62d07m52.41s & 1443 & 4 & 19 & 0.19  & 0.17 & 1.7 & 0.2 \\
I14050-6056 & 7 & 14h08m41.68s & -61d10m46.60s & 2558 & 4 & 32 & 1.21 & 0.68 & 2.2 & 0.5 \\
I14050-6056 & 9 & 14h08m42.79s & -61d10m42.40s & 1844 & 3 & 32 & 0.68 & 0.38 & 3.3 & 0.6 \\
I14206-6151 & 2 & 14h24m23.03s & -62d05m21.32s & 1579 & 3 & 27 & 0.87 & 0.6 & 6.7 & 1.0 \\
I14206-6151 & 5 & 14h24m23.72s & -62d05m12.31s & 2781 & 3 & 27 & 0.96 & 0.66 & 1.4 & 0.4 \\
I14206-6151 & 6 & 14h24m24.11s & -62d05m25.30s & 3263 & 4 & 27 & 0.52 & 0.35 & 0.5 & 0.1 \\
I14212-6131 & 5 & 14h25m03.85s & -61d44m44.95s & 2431 & 5 & 21 & 2.64 & 2.48 & 5.6 & 1.3 \\
I14212-6131 & 11 & 14h24m59.40s & -61d45m04.12s & 3031 & 4 & 21 & 1.28 & 1.21 & 1.4 & 0.4 \\
I15384-5348 & 6 & 15h42m16.81s & -53d58m28.69s & 2969 & 3 & 33 & 1.28 & 0.69 & 1.5 & 0.4 \\
I15384-5348 & 8 & 15h42m17.14s & -53d58m35.86s & 2851 & 4 & 33 & 1.73 & 0.93 & 2.3 & 0.6 \\
I15384-5348 & 9 & 15h42m17.44s & -53d58m35.69s & 2484 & 3 & 33 & 1.45 & 0.78 & 2.9 & 0.7 \\
I15384-5348 & 10 & 15h42m17.31s & -53d58m37.69s & 1937 & 3 & 33 & 0.97 & 0.52 & 4.0 & 0.7 \\
I15384-5348 & 15 & 15h42m17.22s & -53d58m33.80s & 2037 & 4 & 33 & 0.96 & 0.52 & 3.4 & 0.7 \\
I15411-5352 & 15 & 15h44m59.11s & -54d02m13.48s & 3646 & 3 & 30 & 1.51 & 0.91 & 0.9 & 0.3 \\
I15437-5343 & 2 & 15h47m34.01s & -53d52m34.70s & 2665 & 3 & 29 & 2.59 & 1.63 & 4.1 & 1.0 \\
I15437-5343 & 5 & 15h47m32.02s & -53d52m34.12s & 3839 & 4 & 29 & 2.86 & 1.8 & 1.5 & 0.5 \\
\textbf{I15502-5302}$^*$ & \textbf{3} & \textbf{15h54m05.28s} & \textbf{-53d11m40.04s} & \textbf{4894} & \textbf{3} & \textbf{35} & \textbf{20.97} & \textbf{10.55} & \textbf{5.4} & \textbf{2.5} \\
I15522-5411 & 8 & 15h56m09.31s & -54d19m32.17s & 1369 & 6 & 23 & 0.31 & 0.26 & 3.7 & 0.5 \\
I15570-5227 & 8 & 16h00m55.44s & -52d36m27.11s & 4715 & 2 & 28 & 3.64 & 2.39 & 1.1 & 0.5 \\
I15570-5227 & 11 & 16h00m55.17s & -52d36m21.93s & 3535 & 3 & 28 & 1.04 & 0.68 & 0.7 & 0.2 \\
I16026-5035 & 4 & 16h06m25.77s & -50d43m22.44s & 2730 & 1 & 31 & 3.63 & 2.11 & 5.4 & 1.4 \\
I16026-5035 & 6 & 16h06m22.75s & -50d43m35.50s & 3571 & 4 & 31 & 3.02 & 1.75 & 2.0 & 0.7 \\
I16026-5035 & 14 & 16h06m25.95s & -50d43m11.90s & 3169 & 4 & 31 & 1.5 & 0.87 & 1.4 & 0.4 \\
I16037-5223 & 9 & 16h07m38.60s & -52d31m06.77s & 6544 & 4 & 31 & 6.47 & 3.76 & 0.7 & 0.4 \\
I16132-5039 & 5 & 16h17m01.26s & -50d46m47.01s & 2544 & 5 & 32 & 1.66 & 0.93 & 3.1 & 0.7 \\
I16132-5039 & 9 & 16h17m00.73s & -50d46m47.22s & 2154 & 5 & 32 & 0.81 & 0.45 & 2.5 & 0.5 \\
I16132-5039 & 10 & 16h17m02.83s & -50d46m44.45s & 2224 & 4 & 32 & 0.66 & 0.37 & 1.8 & 0.4 \\
I16132-5039 & 12 & 16h17m00.56s & -50d46m46.57s & 2813 & 6 & 32 & 1.25 & 0.7 & 1.7 & 0.4 \\
I16132-5039 & 20 & 16h17m01.82s & -50d46m54.25s & 1683 & 5 & 32 & 0.18 & 0.1 & 1.1 & 0.2 \\
I16272-4837 & 9 & 16h30m57.90s & -48d43m40.47s & 2767 & 5 & 23 & 3.29 & 2.75 & 4.7 & 1.2 \\
I16297-4757 & 3 & 16h33m29.24s & -48d03m26.60s & 3472 & 3 & 27 & 1.24 & 0.85 & 0.9 & 0.3 \\
I16297-4757 & 6 & 16h33m29.82s & -48d03m18.20s & 2698 & 4 & 27 & 1.74 & 1.19 & 2.7 & 0.7 \\
I16297-4757 & 8 & 16h33m29.26s & -48d03m39.46s & 5888 & 2 & 27 & 2.71 & 1.86 & 0.4 & 0.2 \\
I16297-4757 & 9 & 16h33m28.97s & -48d03m39.60s & 5013 & 2 & 27 & 2.08 & 1.43 & 0.5 & 0.2 \\
I16297-4757 & 10 & 16h33m29.20s & -48d03m37.30s & 5801 & 1 & 27 & 2.2 & 1.51 & 0.3 & 0.2 \\
I16304-4710 & 8 & 16h34m05.05s & -47d16m26.87s & 7818 & 4 & 27 & 5.35 & 3.66 & 0.3 & 0.2 \\
I16304-4710 & 10 & 16h34m04.81s & -47d16m24.70s & 10316 & 3 & 27 & 4.53 & 3.11 & 0.1 & 0.1 \\
I16313-4729 & 4 & 16h34m55.59s & -47d35m44.52s & 3268 & 6 & 31 & 4.02 & 2.34 & 3.5 & 1.1 \\
I16313-4729 & 7 & 16h34m55.16s & -47d35m41.47s & 4587 & 3 & 31 & 3.97 & 2.3 & 1.2 & 0.5 \\
I16313-4729 & 8 & 16h34m55.82s & -47d35m46.88s & 2763 & 6 & 31 & 1.95 & 1.13 & 2.8 & 0.7 \\
I16313-4729 & 9 & 16h34m55.94s & -47d35m54.32s & 3241 & 4 & 31 & 4.12 & 2.39 & 3.7 & 1.1 \\
\textbf{I16344-4658}$^*$ & \textbf{4} & \textbf{16h38m09.90s} & \textbf{-47d04m51.68s} & \textbf{7257} & \textbf{3} & \textbf{25} & \textbf{17.3} & \textbf{13.04} & \textbf{1.4} & \textbf{0.9} \\
I16362-4639 & 4 & 16h39m56.62s & -46d45m01.89s & 1516 & 5 & 24 & 0.64 & 0.5 & 5.6 & 0.8 \\
I16362-4639 & 5 & 16h39m58.55s & -46d45m01.63s & 1764 & 4 & 24 & 0.52 & 0.41 & 2.9 & 0.5 \\
I16362-4639 & 7 & 16h39m56.55s & -46d45m03.07s & 1381 & 5 & 24 & 0.51 & 0.4 & 5.9 & 0.8 \\
I16362-4639 & 8 & 16h39m58.67s & -46d45m00.26s & 1597 & 4 & 24 & 0.44 & 0.35 & 3.3 & 0.5 \\
I16362-4639 & 10 & 16h39m58.03s & -46d45m15.10s & 1586 & 5 & 24 & 0.37 & 0.3 & 2.8 & 0.4 \\
I16372-4545 & 9 & 16h40m53.63s & -45d50m54.52s & 3591 & 3 & 26 & 1.79 & 1.29 & 1.2 & 0.4 \\
I16385-4619 & 5 & 16h42m14.73s & -46d25m20.68s & 6354 & 3 & 31 & 11.88 & 6.89 & 1.4 & 0.8 \\
I16424-4531 & 13 & 16h46m05.56s & -45d36m40.54s & 1420 & 6 & 24 & 0.52 & 0.41 & 5.5 & 0.7 \\
I16445-4459 & 6 & 16h48m05.58s & -45d05m14.50s & 6988 & 4 & 24 & 13.49 & 10.69 & 1.2 & 0.8 \\
I16458-4512 & 12 & 16h49m30.68s & -45d18m02.23s & 2387 & 5 & 21 & 4.67 & 4.39 & 10.4 & 2.3 \\
I16489-4431 & 5 & 16h52m34.71s & -44d36m26.79s & 1883 & 3 & 21 & 0.66 & 0.62 & 3.0 & 0.5 \\
I17006-4215 & 8 & 17h04m13.10s & -42d19m54.64s & 1637 & 3 & 27 & 0.44 & 0.3 & 3.0 & 0.5 \\
I17204-3636 & 5 & 17h23m50.76s & -36d38m54.57s & 2187 & 6 & 25 & 1.97 & 1.48 & 5.7 & 1.2 \\
I17204-3636 & 6 & 17h23m50.97s & -36d38m55.04s & 1879 & 5 & 25 & 1.32 & 1.0 & 6.0 & 1.1 \\
I17204-3636 & 10 & 17h23m50.71s & -36d39m04.83s & 2063 & 4 & 25 & 0.58 & 0.44 & 2.0 & 0.4 \\
I17244-3536 & 3 & 17h27m48.68s & -35d39m21.38s & 1437 & 4 & 29 & 1.22 & 0.77 & 12.4 & 1.7 \\
I17244-3536 & 10 & 17h27m49.81s & -35d39m09.70s & 1456 & 3 & 29 & 0.67 & 0.42 & 6.6 & 0.9 \\
I17244-3536 & 12 & 17h27m48.74s & -35d39m05.14s & 2072 & 5 & 29 & 0.78 & 0.49 & 2.7 & 0.5 \\
I17244-3536 & 13 & 17h27m48.20s & -35d39m17.07s & 2247 & 5 & 29 & 0.82 & 0.52 & 2.2 & 0.5 \\
I17244-3536 & 14 & 17h27m49.35s & -35d39m11.02s & 1322 & 4 & 29 & 0.49 & 0.31 & 6.4 & 0.8 \\
I17258-3637 & 11 & 17h29m17.20s & -36d40m15.43s & 894 & 5 & 45 & 0.71 & 0.27 & 30.1 & 2.5 \\
I17269-3312$\_$1 & 3 & 17h30m15.75s & -33d14m50.81s & 3837 & 3 & 22 & 4.22 & 3.74 & 2.3 & 0.8 \\
I17269-3312$\_$1 & 4 & 17h30m15.76s & -33d15m03.13s & 3463 & 3 & 22 & 2.28 & 2.01 & 1.7 & 0.5 \\
I17269-3312$\_$1 & 5 & 17h30m15.34s & -33d14m46.75s & 3340 & 1 & 22 & 2.82 & 2.49 & 2.3 & 0.7 \\
I17269-3312$\_$1 & 9 & 17h30m15.57s & -33d14m46.20s & 2432 & 2 & 22 & 1.0 & 0.88 & 2.1 & 0.5 \\
I17269-3312$\_$2 & 5 & 17h30m14.31s & -33d14m14.28s & 2746 & 2 & 22 & 0.95 & 0.84 & 1.4 & 0.4 \\
I17278-3541 & 5 & 17h31m13.22s & -35d44m11.60s & 657 & 4 & 25 & 0.2 & 0.15 & 21.4 & 1.3 \\
I17278-3541 & 12 & 17h31m13.46s & -35d44m01.58s & 1137 & 3 & 25 & 0.35 & 0.27 & 7.2 & 0.8 \\
I17439-2845 & 9 & 17h47m08.24s & -28d46m11.45s & 3980 & 3 & 30 & 12.2 & 7.37 & 5.9 & 2.2 \\
I17545-2357 & 5 & 17h57m34.47s & -23d57m58.20s & 2219 & 3 & 23 & 1.05 & 0.88 & 2.9 & 0.6 \\
I18075-2040 & 2 & 18h10m34.89s & -20d39m11.25s & 1971 & 3 & 23 & 0.3 & 0.25 & 1.2 & 0.2 \\
I18075-2040 & 4 & 18h10m34.47s & -20d39m09.24s & 2459 & 3 & 23 & 0.16 & 0.13 & 0.3 & 0.1 \\
I18110-1854 & 9 & 18h14m00.97s & -18d53m18.56s & 2311 & 2 & 28 & 1.22 & 0.8 & 3.0 & 0.6 \\
I18116-1646 & 7 & 18h14m35.75s & -16d45m40.22s & 2882 & 6 & 33 & 2.26 & 1.22 & 2.9 & 0.8 \\
I18116-1646 & 10 & 18h14m35.36s & -16d45m34.85s & 5050 & 6 & 33 & 1.66 & 0.89 & 0.4 & 0.2 \\
I18139-1842 & 9 & 18h16m50.89s & -18d41m28.39s & 1987 & 5 & 40 & 1.06 & 0.46 & 4.1 & 0.8 \\
I18223-1243 & 9 & 18h25m10.76s & -12d42m20.10s & 3059 & 6 & 23 & 1.25 & 1.05 & 1.3 & 0.4 \\
I18290-0924 & 3 & 18h31m42.68s & -9d22m27.48s & 2666 & 5 & 22 & 6.0 & 5.31 & 9.6 & 2.4 \\
I18290-0924 & 5 & 18h31m43.07s & -9d22m23.06s & 3090 & 5 & 22 & 1.91 & 1.69 & 2.0 & 0.6 \\
I18290-0924 & 6 & 18h31m43.72s & -9d22m12.20s & 3387 & 4 & 22 & 1.59 & 1.41 & 1.2 & 0.4 \\
I18290-0924 & 7 & 18h31m43.55s & -9d22m30.92s & 2656 & 3 & 22 & 0.97 & 0.86 & 1.6 & 0.4 \\
I18290-0924 & 9 & 18h31m43.26s & -9d22m33.82s & 3503 & 4 & 22 & 0.92 & 0.82 & 0.6 & 0.2 \\
I18290-0924 & 11 & 18h31m43.11s & -9d22m13.70s & 4469 & 1 & 22 & 1.01 & 0.9 & 0.3 & 0.1 \\
I18290-0924 & 12 & 18h31m43.50s & -9d22m29.54s & 2683 & 4 & 22 & 0.68 & 0.6 & 1.1 & 0.3 \\
I18308-0503 & 2 & 18h33m30.45s & -5d00m55.16s & 2546 & 5 & 31 & 2.24 & 1.3 & 4.1 & 1.0 \\
I18308-0503 & 4 & 18h33m29.22s & -5d01m00.13s & 3230 & 1 & 31 & 1.23 & 0.72 & 1.1 & 0.3 \\
I18308-0503 & 7 & 18h33m29.38s & -5d00m59.50s & 2685 & 4 & 31 & 0.81 & 0.47 & 1.3 & 0.3 \\
I18308-0503 & 8 & 18h33m29.59s & -5d00m57.60s & 1953 & 3 & 31 & 0.37 & 0.21 & 1.5 & 0.3 \\
I18314-0720 & 16 & 18h34m08.48s & -7d17m50.41s & 3488 & 3 & 30 & 1.77 & 1.07 & 1.3 & 0.4 \\
I18317-0513 & 8 & 18h34m26.36s & -5d10m56.10s & 2783 & 2 & 31 & 1.91 & 1.11 & 2.7 & 0.7 \\
I18317-0513 & 14 & 18h34m25.49s & -5d11m00.10s & 1518 & 3 & 31 & 0.38 & 0.22 & 3.3 & 0.5 \\
I18317-0757 & 5 & 18h34m25.29s & -7d54m44.68s & 2386 & 2 & 30 & 2.69 & 1.62 & 6.0 & 1.3 \\
I18317-0757 & 12 & 18h34m25.36s & -7d54m41.40s & 5002 & 1 & 30 & 5.7 & 3.44 & 1.4 & 0.6 \\
I18411-0338 & 8 & 18h43m46.23s & -3d35m24.21s & 5616 & 5 & 27 & 4.38 & 3.0 & 0.7 & 0.4 \\
I18440-0148 & 2 & 18h46m36.14s & -1d45m16.98s & 9251 & 3 & 33 & 4.8 & 2.59 & 0.2 & 0.2 \\
I18440-0148 & 4 & 18h46m36.91s & -1d45m28.32s & 5407 & 4 & 33 & 1.94 & 1.05 & 0.4 & 0.2 \\
I18440-0148 & 6 & 18h46m36.33s & -1d45m19.13s & 9054 & 4 & 33 & 3.19 & 1.72 & 0.1 & 0.1 \\
I18440-0148 & 7 & 18h46m37.17s & -1d45m17.23s & 7086 & 3 & 33 & 2.47 & 1.33 & 0.2 & 0.1 \\
I18445-0222 & 3 & 18h47m09.50s & -2d18m48.17s & 6251 & 3 & 27 & 9.35 & 6.41 & 1.2 & 0.7 \\
I18445-0222 & 6 & 18h47m09.62s & -2d18m42.45s & 4642 & 5 & 27 & 3.21 & 2.2 & 1.0 & 0.4 \\
I18445-0222 & 7 & 18h47m09.73s & -2d18m42.22s & 4590 & 6 & 27 & 3.06 & 2.1 & 1.0 & 0.4 \\
I18445-0222 & 9 & 18h47m10.72s & -2d18m46.60s & 3262 & 3 & 27 & 1.39 & 0.95 & 1.2 & 0.4 \\
I18461-0113 & 5 & 18h48m43.03s & -1d10m03.78s & 3170 & 1 & 27 & 5.23 & 3.59 & 5.0 & 1.5 \\
I18507+0121 & 2 & 18h53m18.28s & 1d25m13.68s & 2651 & 3 & 22 & 7.0 & 6.19 & 11.4 & 2.8 \\
I18507+0121 & 7 & 18h53m18.75s & 1d25m26.47s & 2455 & 3 & 22 & 3.13 & 2.77 & 6.4 & 1.5 \\
I18530+0215 & 7 & 18h55m32.97s & 2d19m02.74s & 4296 & 5 & 26 & 7.45 & 5.35 & 2.8 & 1.1 \\
I19097+0847 & 8 & 19h12m09.62s & 8d52m06.24s & 6281 & 4 & 23 & 5.74 & 4.8 & 0.7 & 0.4 \\
I19097+0847 & 11 & 19h12m09.60s & 8d52m13.00s & 3307 & 5 & 23 & 3.03 & 2.53 & 2.5 & 0.8 \\
\enddata
\begin{flushleft}
$^{(a)}$The number of molecular lines detected for each core.\\
$^{(b)}$ The dust temperature of the parental clump.\\
$^{(c)}$ The upper and lower masses were calculated corresponding to the assumed 20\,K typical for candidate starless cores in the IR-bright environment and the average temperature of the parental clump, respectively. Note that since there are 2 cores (i.e., in I13295-6152 clumps) whose parental clump has an average temperature slightly lower than 20\,K, the upper and lower mass estimates were switched manually. \\
$^{(d)}$ The number and the surface density of each core are estimated for the 20\,K temperature.\\
$^*$ Candidate high-mass starless core.
\end{flushleft}

\end{deluxetable*}

\startlongtable
\begin{deluxetable}{ccccccccccc}\label{table:phywarm}
% \tabletypesize{\scriptsize}
% \tablewidth{0pt} 
% \tablenum{4}
\tablecaption{Derived parameters for candidate warm and evolved cores.} 
% \tablecomments{}
% \centering
\tablehead{
\colhead{Source} & \colhead{Core} & \colhead{RA\,(IRCS)} & \colhead{Dec\,(IRCS)}& \colhead{$R$}& \colhead{$N_{\rm line}^{(\it {a})}$}& \colhead{$\rm Temp.^{(\it b)}$}& \colhead{Mass}& \colhead{$n_{\rm H_2}$}& \colhead{$\Sigma$}& \colhead{$\rm Evol.^{(\it c)}$}\\
&& \colhead{(h:m:s)}  & \colhead{(d:m:s)}& \colhead{(au)} & &(K)& \colhead{(\msun)} & $(\rm\times 10^5\,cm^{-3})$ & $(\rm g\,cm^{-2})$& \\ }
\colnumbers 
\startdata 
I08303-4303 & 1 & 8h32m08.64s & -43d13m45.65s & 1180 & 48 & 100.0 & 2.54 & 468.92 & 5.17 & HMC \\
I08303-4303 & 2 & 8h32m09.05s & -43d13m43.33s & 949 & 11 & 30.0 & 2.41 & 854.36 & 7.57 & WC \\
I08303-4303 & 3 & 8h32m08.46s & -43d13m49.13s & 1374 & 18 & 100.0 & 1.0 & 116.99 & 1.5 & HMC \\
I08303-4303 & 4 & 8h32m08.46s & -43d13m47.22s & 1157 & 16 & 100.0 & 0.54 & 105.47 & 1.14 & HMC \\
I08303-4303 & 5 & 8h32m08.90s & -43d13m52.12s & 1094 & 10 & 30.0 & 0.45 & 103.52 & 1.06 & WC \\
I08303-4303 & 6 & 8h32m08.88s & -43d13m42.05s & 2006 & 11 & 30.0 & 0.55 & 20.63 & 0.39 & WC \\
I08448-4343 & 1 & 8h46m32.31s & -43d54m36.85s & 478 & 14 & 30.0 & 1.18 & 3278.14 & 14.64 & WC \\
I08448-4343 & 2 & 8h46m33.39s & -43d54m37.28s & 583 & 15 & 30.0 & 0.93 & 1427.41 & 7.77 & WC \\
I08448-4343 & 3 & 8h46m34.84s & -43d54m30.10s & 401 & 8 & 100.0 & 0.08 & 377.86 & 1.42 & UCH{\sc ii}-{\it l} \\
I08448-4343 & 4 & 8h46m34.32s & -43d54m39.51s & 462 & 7 & 30.0 & 0.19 & 586.74 & 2.53 & WC \\
I08448-4343 & 5 & 8h46m31.17s & -43d54m36.47s & 451 & 5 & 30.0 & 0.2 & 663.82 & 2.8 & WC \\
I08448-4343 & 6 & 8h46m34.94s & -43d54m23.65s & 534 & 10 & 30.0 & 0.25 & 488.04 & 2.43 & WC \\
I08448-4343 & 7 & 8h46m34.63s & -43d54m32.50s & 711 & 9 & 30.0 & 0.33 & 275.41 & 1.83 & WC \\
I08448-4343 & 8 & 8h46m31.79s & -43d54m36.02s & 470 & 9 & 30.0 & 0.16 & 454.3 & 1.99 & WC \\
I08448-4343 & 10 & 8h46m33.33s & -43d54m38.56s & 504 & 10 & 30.0 & 0.19 & 458.13 & 2.16 & WC \\
I08448-4343 & 11 & 8h46m33.30s & -43d54m34.67s & 439 & 10 & 30.0 & 0.17 & 604.03 & 2.48 & WC \\
I08448-4343 & 12 & 8h46m33.65s & -43d54m34.66s & 461 & 11 & 30.0 & 0.15 & 473.6 & 2.04 & WC \\
I08448-4343 & 13 & 8h46m35.12s & -43d54m24.25s & 630 & 13 & 30.0 & 0.27 & 331.56 & 1.95 & WC \\
I08448-4343 & 14 & 8h46m34.90s & -43d54m26.72s & 428 & 11 & 30.0 & 0.11 & 438.45 & 1.75 & WC \\
I08448-4343 & 15 & 8h46m33.74s & -43d54m34.22s & 481 & 10 & 30.0 & 0.13 & 360.88 & 1.62 & WC \\
I08448-4343 & 16 & 8h46m32.22s & -43d54m40.99s & 488 & 6 & 30.0 & 0.1 & 263.18 & 1.2 & WC \\
I08448-4343 & 17 & 8h46m32.96s & -43d54m34.82s & 437 & 9 & 30.0 & 0.1 & 370.67 & 1.51 & WC \\
I08448-4343 & 18 & 8h46m33.57s & -43d54m34.90s & 376 & 9 & 30.0 & 0.09 & 484.54 & 1.7 & WC \\
I08448-4343 & 19 & 8h46m32.13s & -43d54m41.71s & 382 & 6 & 30.0 & 0.05 & 297.98 & 1.06 & WC \\
I08448-4343 & 21 & 8h46m32.50s & -43d54m37.09s & 517 & 11 & 30.0 & 0.09 & 194.24 & 0.94 & WC \\
I08448-4343 & 22 & 8h46m34.74s & -43d54m20.61s & 1013 & 7 & 30.0 & 0.15 & 44.31 & 0.42 & WC \\
I08448-4343 & 23 & 8h46m32.66s & -43d54m35.87s & 1583 & 8 & 30.0 & 0.69 & 52.48 & 0.78 & WC \\
I08448-4343 & 24 & 8h46m33.37s & -43d54m28.29s & 1201 & 9 & 30.0 & 0.09 & 15.09 & 0.17 & WC \\
I08448-4343 & 25 & 8h46m34.83s & -43d54m32.67s & 670 & 9 & 30.0 & 0.07 & 69.81 & 0.44 & WC \\
I08448-4343 & 26 & 8h46m33.76s & -43d54m37.19s & 945 & 7 & 30.0 & 0.07 & 25.4 & 0.22 & WC \\
\enddata
\begin{flushleft}
(The complete table is available in machine-readable form.)\\
$^{(a)}$ The number of molecular lines detected for each core.\\
$^{(b)}$ The dust temperature for the warm and evolved core candidates was assumed to be 30\,K and 100\,K, respectively. \\
$^{(c)}$ The evolutionary stage of each core: 'WC', 'HMC', 'UC\hii-{\it h}' and UC\hii-{\it l}, corresponding to candidates of warm core, hot molecular core, UC\hii core  with a high line richness ($N_{\rm line}\ge15$), and UC\hii core with a low line richness ($6<N_{\rm line}<15$), respectively. \\
\end{flushleft}
\end{deluxetable}
% \clearpage

\section{Impact of various clump distances}
\begin{figure*}[ht!]
    \centering
    \includegraphics[angle=0, width=0.5\textwidth]{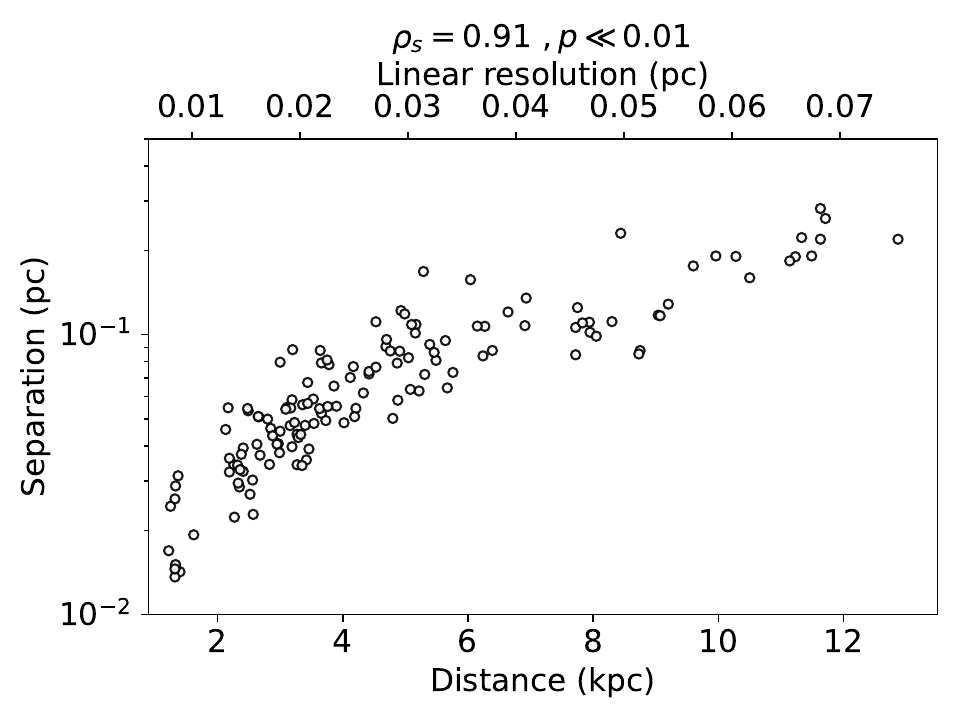} 

    \caption{Distribution of the median linear core separation within each clump against the source distance. 
    The Spearman’s rank correlation coefficient ($\rho_s=0.9$) and the probability value ($p\ll0.01$) of the distribution are shown on the top.
    On the upper x-axis also displays the spatial resolution at the corresponding distance.}
    \label{fig:msep.dis}   
\end{figure*}

\begin{figure}[ht!]
    \centering
    \includegraphics[angle=0, width=0.47\textwidth]{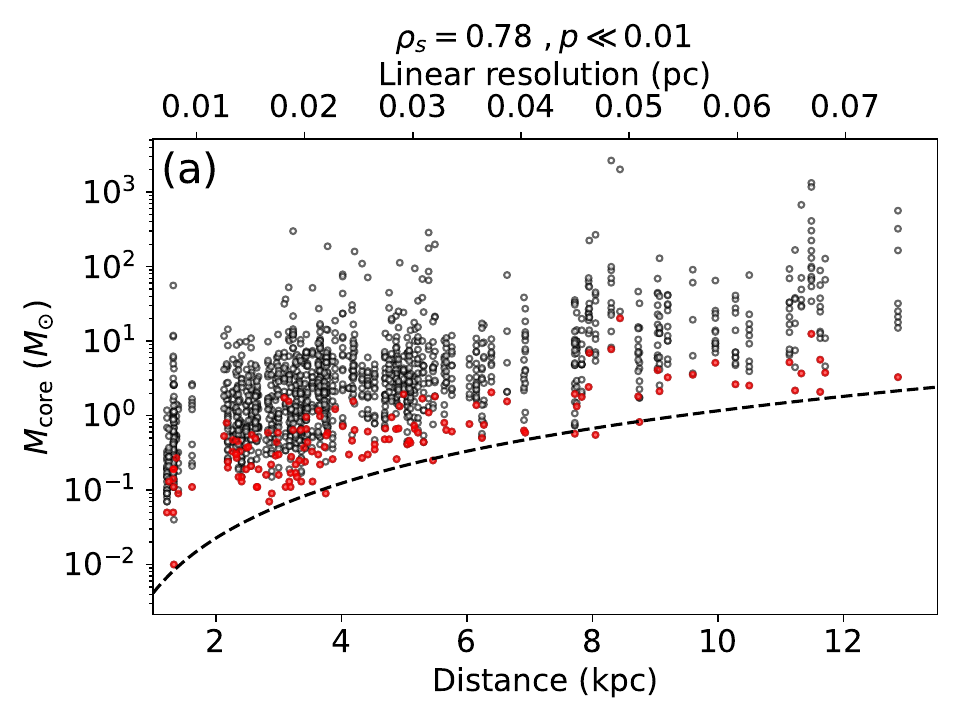} 
    \includegraphics[angle=0, width=0.47\textwidth]{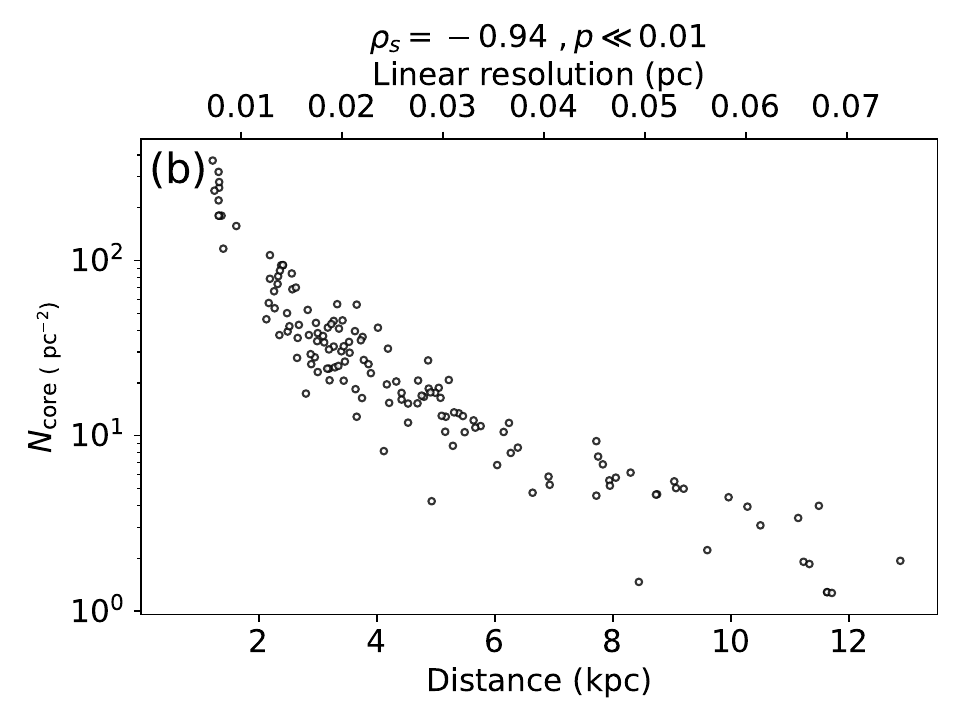} 
    \caption{Panel\,(a): distribution of the core mass against the source distance. The red circle symbols display the minimum core mass detected within each clump. The black dashed curve fits the lower envelope of the minimum core masses as a function of the distance. 
    Panel\,(b): distribution of the surface density of the core count within each clump against the source distance. 
    On the top of both panels shows the Spearman’s rank correlation coefficient ($\rho_s$) and the probability value ($p$) of the distribution.
    On the upper x-axis of both panels also displays the spatial resolution at the corresponding distance.}
    \label{fig:m.dis}   
\end{figure}

\begin{figure}[ht!]
    \centering
    \includegraphics[angle=0, width=0.47\textwidth]{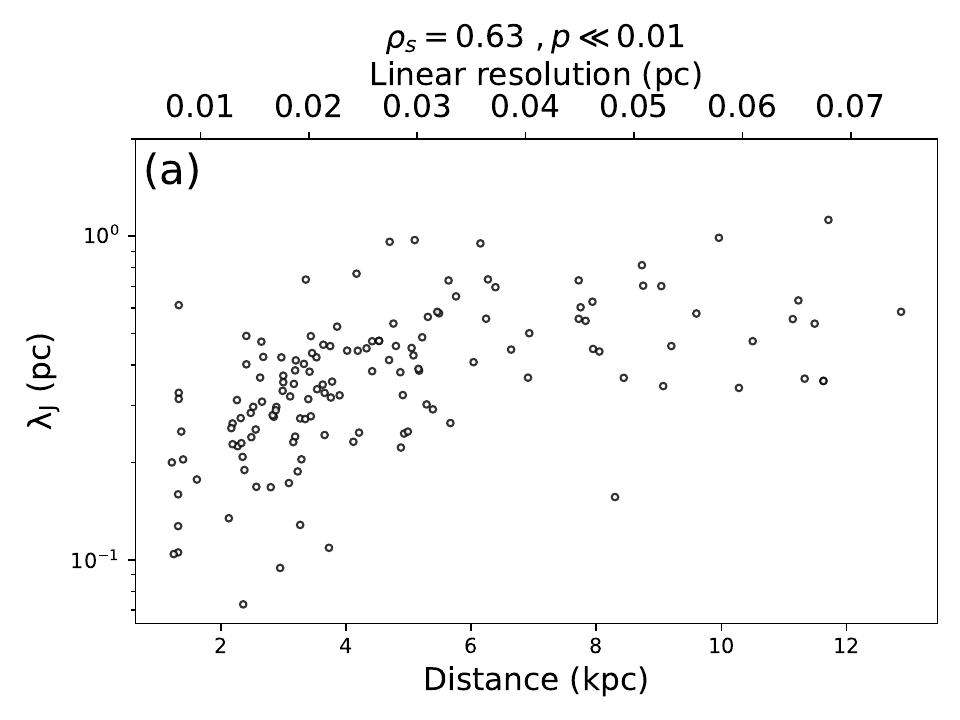} 
    \includegraphics[angle=0, width=0.47\textwidth]{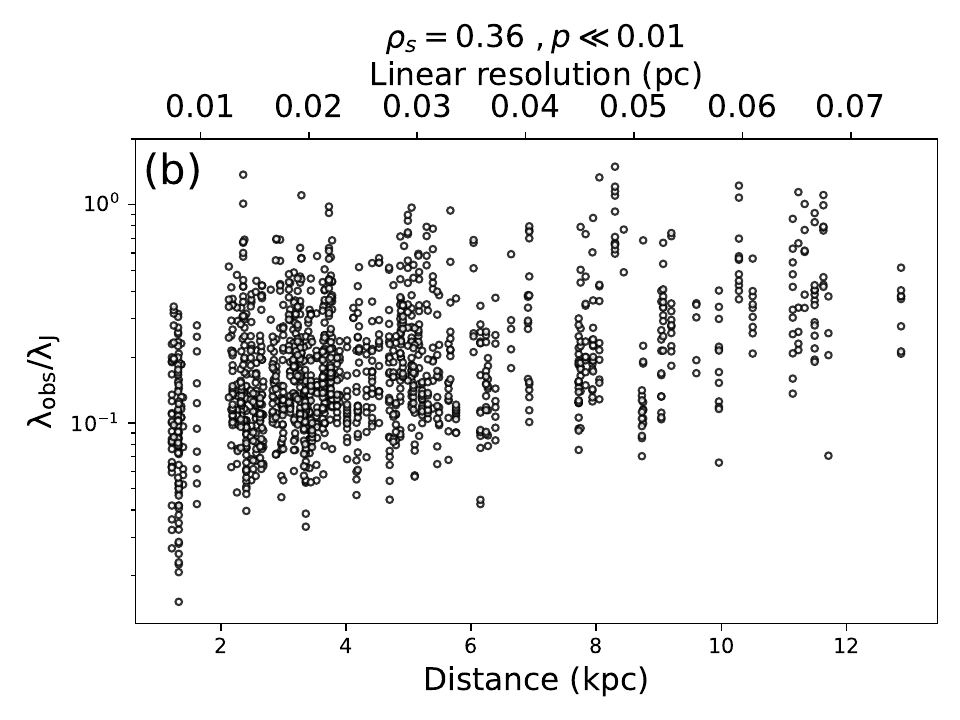} 
    \caption{Panle\,(a) Thermal Jeans length vs. protocluster clumps distance. Panel\,(b): Ratio of core separation to thermal Jeans length vs. protocluster clumps distance.
    } 
    \label{fig:s_j.dis}   
\end{figure}

Figure\,\ref{fig:msep.dis} shows the distribution of median linear core separation within each clump as a function of source distance, revealing a strong positive correlation that corresponds to a Spearman's coefficient of  $\rho_s \sim 0.9$ and a $p$ value of $\ll 0.01$. Figure\,\ref{fig:m.dis} displays the relationships between source distance and core mass (panel\,a) and the surface density of core count within each clump (panel\,b). Core mass shows a strong positive correlation with distance ($\rho_s \sim 0.8$ and $p \ll 0.01$), while core count surface density does a strong negative correlation ($\rho_s \sim -0.9$ and $p \ll 0.01$). Figure\,\ref{fig:s_j.dis}\,a plots the thermal Jeans length of the clump against source distance, showing a positive correlation ($\rho_s \sim 0.6$ and $p \ll 0.01$). Figure\,\ref{fig:s_j.dis}\,b presents the ratio $\lambda_{\rm obs}/\lambda_{\rm J}$ as a function of the source distance, indicating a weak or no correlation ($\rho_s \sim 0.4$ and $p \ll 0.01$) between these two quantities.

\section{Additional information of candidate starless cores}
\begin{figure*}[ht!]
    \centering
    \includegraphics[angle=0, width=1\textwidth]{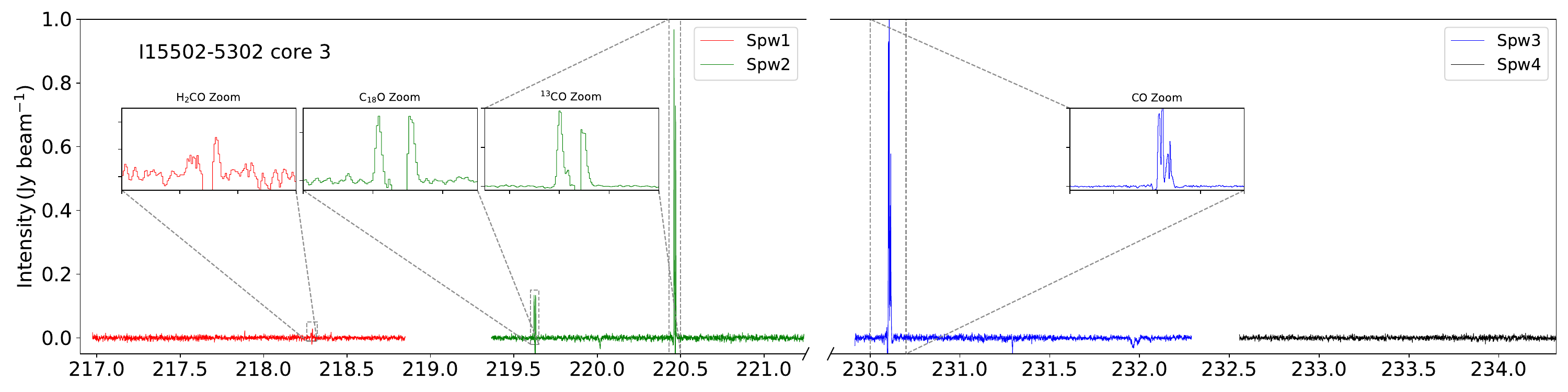} 
    \includegraphics[angle=0, width=1\textwidth]{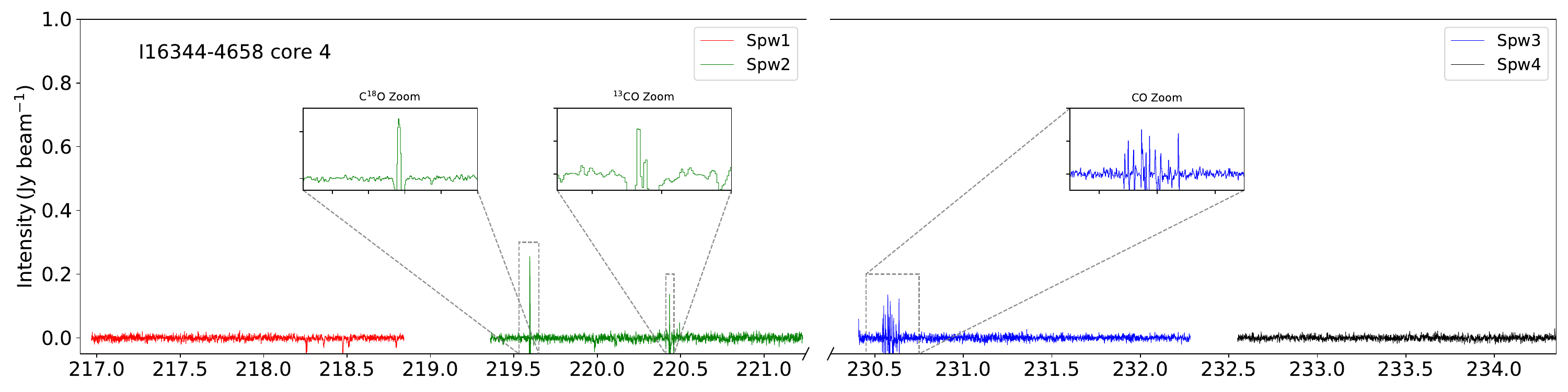} 
    \caption{Core-averaged spectra of two high-mass starless core candidates from four QUARKS SPWs. Note that the zoom-in of the CO line spectrum in the bottom panel presents many artificial absorption dips caused by missing flux of extended gas structures, which however do not affect our classification of candidate starless cores. These artifacts can be addressed in the future by combing our QUARKS and single-dish observations.}
    \label{fig:pscline}    
\end{figure*}

\begin{figure*}[ht!]
    \centering
    \includegraphics[angle=0, width=0.45\textwidth]{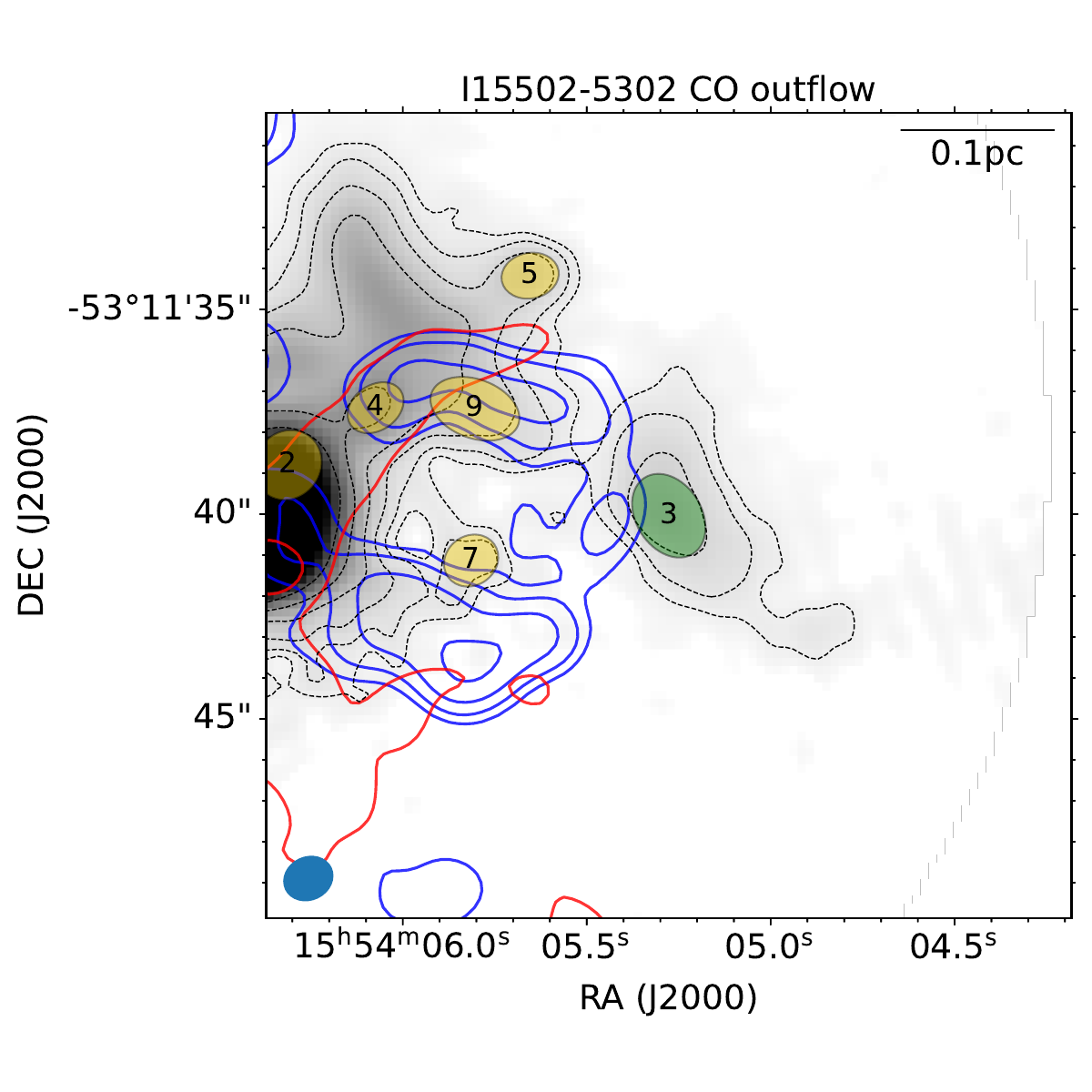} 
    \includegraphics[angle=0, width=0.46\textwidth]{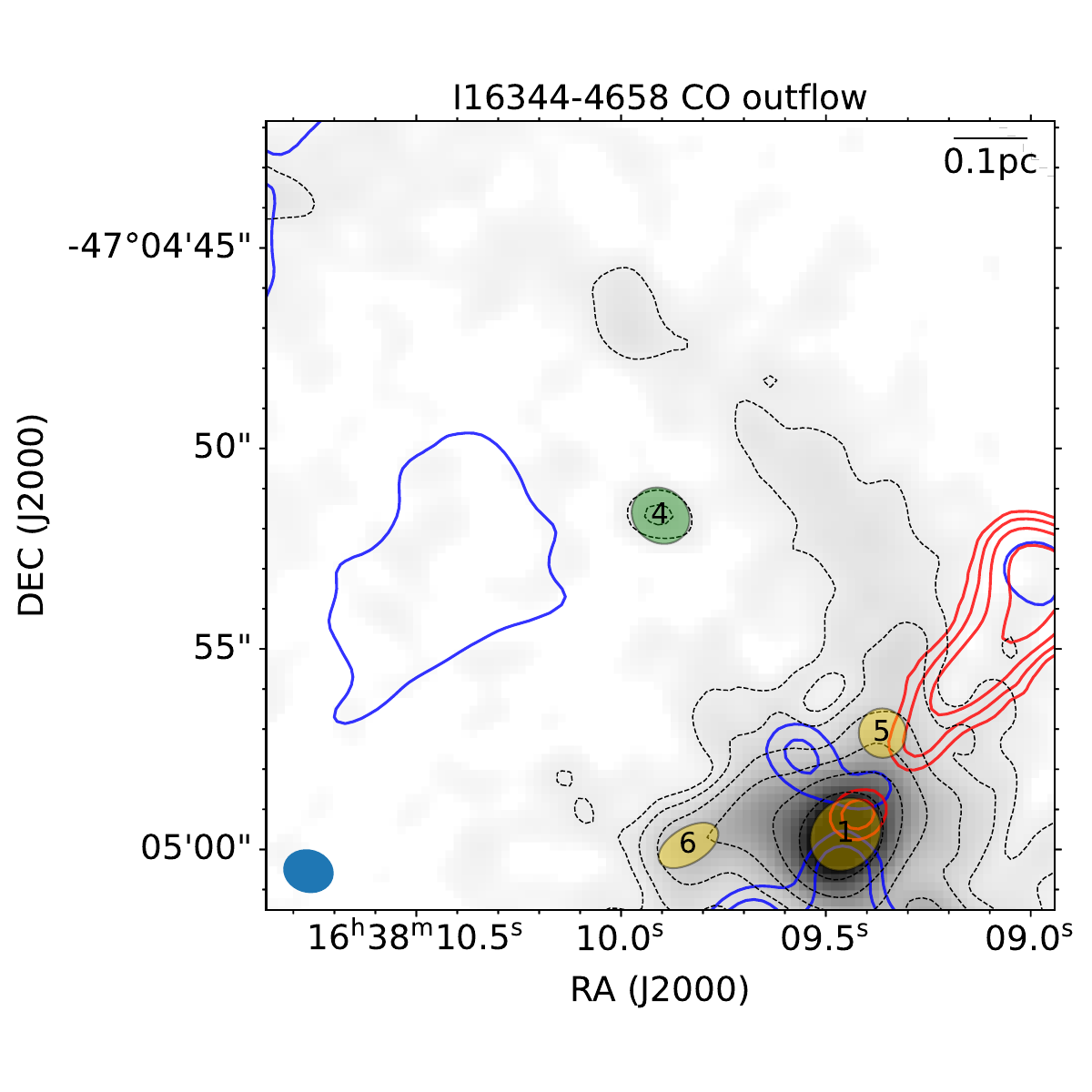} 
    \caption{CO molecular outflow overlaid on the 1.3\,mm dust continuum map of two protocluster clumps containing high-mass starless core candidates. The velocity offset is set around $\rm\,V_{LSR}\,\pm\,10-30\,km\,s^{-1}$. 
    The red and blue contours start at 3\,rms ($\rm\sim\,0.2\,mJy\,beam^{-1}\,km\,s^{-1}$) and follow as [6, 12, 24, 48]\,rms. High-mass starless core candidates in this study are shown as green circles. ALMA-QUARKS TM2+ACA 1.3\,mm dust continuum contour levels are [3, 6, 12, 24, 48]\,rms, with rms is of $\rm\sim 0.6\,mJy\,beam^{-1}$.} 
    \label{fig:cooutflow}    
\end{figure*}

\begin{figure}[ht!]
    \centering
    \includegraphics[angle=0, width=0.5\textwidth]{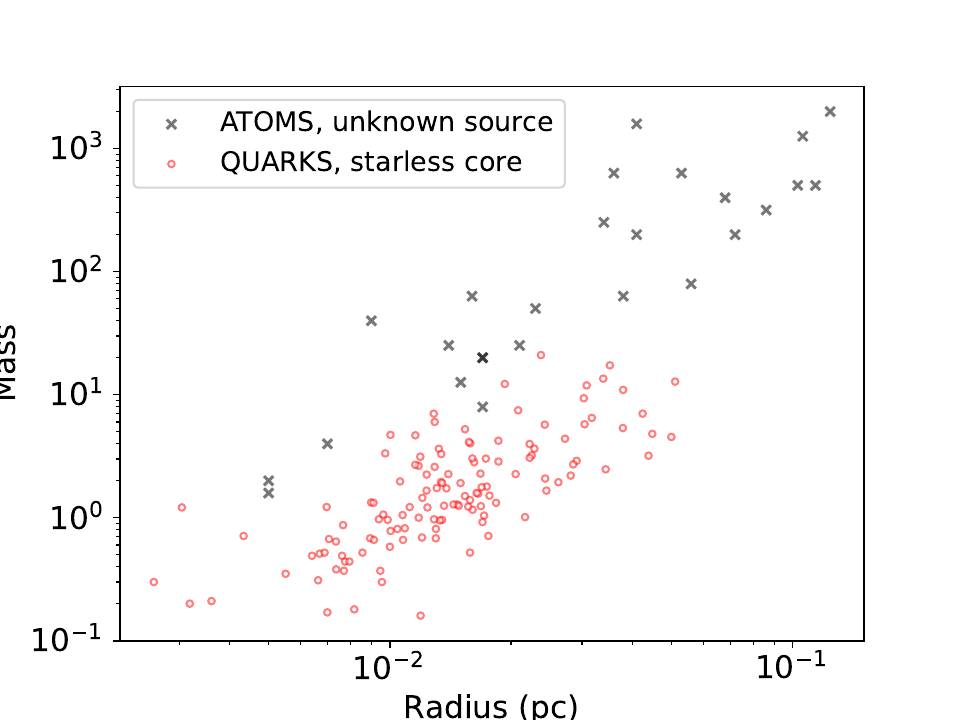} 
    \caption{Mass-Radius relation between ATOMS ``unknown" source and candidate starless core of the QUARKS TM2+ACA catalogue.
    } 
    \label{fig:atomsmass:quarksmass}   
\end{figure}

We provide here additional information for two candidate high-mass starless cores. The molecular line emission of each high-mass starless core candidate is shown in Figure\,\ref{fig:pscline}, and the CO molecular outflow maps overlaid on the QUARKS 1.3\,mm continuum image are presented in Figure\,\ref{fig:cooutflow}.
In Figure\,\ref{fig:atomsmass:quarksmass}, we present a comparison  between ATOMS ``unknown" sources \citep{2021MNRAS.505.2801L} and QUARKS TM2+ACA starless cores for the mass and size parameters. Note that only 26 ATOMS ``unknown" sources that are related to candidate starless cores from the QUARKS survey are depicted in this figure. We find that the ATOMS ``unknown" sources generally exhibit about 1dex larger sizes and masses. This suggests that the lower angular resolution at 3\,mm is capable of detecting more extended envelope emissions, while the higher angular resolution at 1.3\,mm is better suited to trace the kernel of the cores.

% \begin{figure*}[ht!]
%     \centering
%     \includegraphics[angle=0, width=0.9\textwidth]{15502_3_example_v1.pdf} 
%     \includegraphics[angle=0, width=0.9\textwidth]{16344_4_example_v1.pdf} 
%     \caption{Core-averaged spectra of two high-mass starless core candidates from four QUARKS SPWs.} 
%     \label{fig:pscline}    
% \end{figure*}

\clearpage

\bibliographystyle{aasjournal}
\bibliography{reference}{}

\end{document}